\documentclass[sigconf]{acmart}

\AtBeginDocument{%
  \providecommand\BibTeX{{%
    \normalfont B\kern-0.5em{\scshape i\kern-0.25em b}\kern-0.8em\TeX}}}

\usepackage{amsmath,amsfonts}
\usepackage{algorithmic}
\usepackage{graphicx}
\usepackage{textcomp}
\usepackage{xcolor}
\usepackage{caption}
\usepackage{listings}
\usepackage{soul}
\usepackage{here}

\newcommand{\itbold}[1]{\textbf{\textit{#1}}}
\newcommand{\pdif}[2]{\frac{\partial #1}{\partial #2}}


\copyrightyear{2021}
\acmYear{2021}
\setcopyright{acmcopyright}\acmConference[SC '21]{The International Conference for
High Performance Computing, Networking, Storage and Analysis}{November 14--19,
2021}{St. Louis, MO, USA}
\acmBooktitle{The International Conference for High Performance Computing,
Networking, Storage and Analysis (SC '21), November 14--19, 2021, St. Louis, MO,
USA}
\acmPrice{15.00}
\acmDOI{10.1145/3458817.3487401}
\acmISBN{978-1-4503-8442-1/21/11}

\begin{document}


\title[400 Trillion-Grid Vlasov Simulation of Comic Relic Neutrinos]{A 400 Trillion-Grid Vlasov Simulation on Fugaku Supercomputer: 
Large-Scale Distribution of Cosmic Relic Neutrinos in a Six-dimensional Phase Space}

\author{Kohji Yoshikawa}
\email{kohji@ccs.tsukuba.ac.jp}
\authornote{corresponding author}
\affiliation{%
\institution{Center for Compuational Sciences, University of Tsukuba}
\city{Tsukuba}
\country{Japan}
}

\author{Satoshi Tanaka}
\email{satoshi.tanaka@yukawa.kyoto-u.ac.jp}
\affiliation{%
\institution{Yukawa Institute for Theoretical Physics, Kyoto University}
\city{Kyoto}
\country{Japan}
}

\author{Naoki Yoshida}
\email{naoki.yoshida@ipmu.jp}
\affiliation{
\institution{Kavli Institute for the Physics and Mathematics of the Universe, The University of Tokyo}
\city{Kashiwa}
\country{Japan}
}

\begin{abstract}
 We report a Vlasov simulation of cosmic relic neutrinos combined with 
 {\boldmath $N$}-body simulation of cold dark matter in the context of 
 large-scale structure formation in the Universe performed on Fugaku
 supercomputer.  Gravitational dynamics of the neutrinos is followed, 
 for the first time, by directly integrating the Vlasov equation in 
 a six-dimensional phase space. Our largest simulation combines the 
 Vlasov simulation on 400 trillion grids and 330 billion-body calculations 
 in a self-consistent manner, and reproduces accurately the nonlinear 
 dynamics of neutrinos in the Universe. The novel high-order Vlasov 
 solver is optimized by combining an array of state-of-the-art numerical 
 schemes and fully utilizing the SIMD instructions on the A64FX processors. 
 Time-To-Solution of our simulation is an order of magnitude shorter than 
 the largest {\boldmath $N$}-body simulations. The performance scales excellently 
 with up to 147,456 nodes (7 million CPU cores) on Fugaku; the weak and strong scaling efficiencies are 82\% -- 96\% and 82\% -- 93\%, respectively.
\end{abstract}

\maketitle

\section{Justification for ACM Gordon Bell Prize}

We present a series of hybrid Vlasov/$N$-body simulation of the
large-scale structure formation in the Universe. This includes the
world's first Vlasov simulation of cosmic relic neutrinos performed on a
full six-dimensional phase space domain, and the largest Vlasov
simulation ever conducted. This is also the first successful run in the
world that combines the complementary advantages of the particle-based
$N$-body simulation and the Vlasov simulation for a mixture of different
kinds of matter components. Our simulations are performed on Fugaku
supercomputer installed at RIKEN Center for Computational Sciences with
up to 147,456 nodes (7,077,888 CPU cores). We achieve very high
scalability of Vlasov simulations and also of the whole end-to-end
simulations both for weak and strong scaling efficiencies. At the same
time, the time-to-solution is improved by an order of magnitude to
obtain numerical results on the dynamics of massive neutrinos in the
Universe with an equivalent spatial resolution and with much superior
discreteness noise level to those of existing state-of-the-art
particle-based $N$-body simulations.

\section{Performance Attributes}

\begin{table}[H]
 \normalsize
 \begin{tabular}[t]{ll}
  \toprule
  Category of achievement & scalability, time-to-solution \\
  Type of method used & explicit \\
  Results reported on the basis of & whole application including I/O \\
  Precision reported & mixed precision \\
  System scale & measured on the full system \\
  Measurement mechanism & timers \\
  \bottomrule
 \end{tabular}
\end{table}

\section{Overview of the Problem}

Neutrinos are elementally particles that are assumed to be massless like
photons in the standard model of particle physics. Discovery of neutrino
oscillation \cite{Fukuda1998} revealed, however, that neutrinos have
finite masses, suggesting some unknown physics beyond the standard
model.  Despite its fundamental importance in understanding the origin
of matter and anti-matter asymmetry, the absolute mass-scale of
neutrinos remains highly uncertain. So far, the neutrino oscillation 
experiments provide only {\it lower bounds} on the neutrino mass. Although 
several terrestrial particle experiments have been conducted to measure the neutrino 
mass through tritium beta decay and neutrinoless 
double beta decay, such experiments place only upper limits 
on the total absolute mass of neutrinos.

A promising approach is to measure the neutrino mass through the
dynamical effect on cosmic structure formation.  The standard
cosmological model posits that the large-scale structure (LSS) of the
Universe formed through gravitational amplification of tiny density
fluctuations left over from the Big Bang.  The model also predicts that
there exist 'relic' neutrinos that permeate the Universe from the early
through to the present epoch.  The fractional energy density of the
massive neutrinos scales with the total mass of the three neutrino
species, and is estimated to be of the order $10^{-3}$--$10^{-2}$. Despite the
small contribution to the present-day cosmic energy budget, relic neutrinos with 
finite mass (hereafter massive neutrinos) influence significantly the LSS 
formation through gravitational interaction with other non-relativistic matter 
dominated by the so-called cold dark matter (CDM). The primary effect of 
massive neutrinos is to suppress the nonlinear growth of large-scale 
density fluctuations through collisionless damping. 
The massive neutrinos have very large velocity
dispersion and effectively prevent clustering of themselves and of other
matter. Since the velocity dispersion directly depends on the neutrino mass, 
we can, in principle, constrain or measure the neutrino mass by detecting 
and precisely modeling the collisionless damping effect imprinted in the LSS. 
This offers a novel and promising method for {\it measuring} the neutrino mass 
from cosmological observations such as galaxy surveys. 

So far, particle-based $N$-body methods have been a primary choice in numerical
simulations of the cosmic structure formation. The gravitational dynamics of
CDM and massive neutrinos are followed numerically by $N$-body methods
with employing a large number of particles 
\cite{Bird2012, Inman2015, Inman2017, Banerjee2018}. Unfortunately, 
there remain several intrinsic drawbacks in such $N$-body simulations. 
An $N$-body simulation statistically samples the matter distribution in 
the six-dimensional phase space (three-dimensional physical space and
three-dimensional velocity or momentum space) using a finite number of 
discrete ``super-particles'' in a Monte-Carlo manner. The numerical results are then susceptible 
to the well-known shot noise. The discreteness noise critically compromises 
the results when a "hot" component with a very large velocity dispersion 
like massive neutrinos is simulated (see our results in \S\ref{sec:result}). 
Furthermore, particle-based methods are not well-suited to accurately
reproduce collisionless damping, in which the high-velocity component
in the tail of the velocity distribution plays a crucial role.  Clearly, 
it is desirable to adopt a numerical scheme that accurately represents the 
continuous and extended velocity distribution in a multi-dimensional phase 
space.

Here, we propose a completely new approach that explicitly follows the dynamics of massive neutrinos by solving time evolution of their distribution function with the finite volume method. Our approach 
eliminates the above-described numerical problems by representing the massive 
neutrino as a continuous medium in the six-dimensional phase space. This
approach enables us to reproduce the neutrino distribution {\it without} shot
noise, even when the velocity distribution has a broad, extended tail 
\cite{Yoshikawa2013}. 

Since the cosmic relic neutrinos can be regarded as a collisionless matter, the
time evolution of their distribution function is described by
the collisionless Boltzmann equation or the Vlasov equation:
\begin{equation}
 \label{eq:vlasov}
 \begin{split}
  \pdif{f(\itbold{x}, \itbold{u}, t)}{t} + \frac{\itbold{u}}{a(t)^2}&\cdot\pdif{f(\itbold{x}, \itbold{u}, t)}{\itbold{x}}\\
  & - \pdif{\phi(\itbold{x},t)}{\itbold{x}}\cdot\pdif{f(\itbold{x}, \itbold{u}, t)}{\itbold{u}}=0,
 \end{split}
\end{equation}
where $a(t)$ is the scale factor describing the time dependence of the
cosmic expansion, $f(\itbold{x}, \itbold{u}, t)$ is the distribution
function of massive neutrinos as a function of the comoving spatial
position $\itbold{x}$ and the canonical velocity
$\itbold{u}=a(t)^2\dot{\itbold{x}}$.  The gravitational potential
$\phi(\itbold{x})$ satisfies the Poisson equation given by
\begin{equation}
 \label{eq:poisson}
 \nabla^2 \phi(\itbold{x},t) = 4\pi G a(t)^2 [\rho(\itbold{x},t)-\bar{\rho}(t)],
\end{equation}
where $G$ is the gravitational constant, and $\rho(\itbold{x},t)$ and
$\bar{\rho}(t)$ are the mass density field and its spatial average,
respectively. Hereafter, our approach that directly integrates equations
(1) and (2) is referred to as Vlasov simulation. 

\section{Current State of the Art}

The currently largest $N$-body simulation of the LSS
in the Universe that includes massive neutrinos is
the TianNu simulation performed on China's Tianhe-2 supercomputer employing 
$6912^3$ CDM particles
and $13824^3$ neutrino particles \cite{Yu2017,Emberson2017}.
Their numerical code, CUBEP$^3$M, adopts a variant of 
Particle--Particle--Particle--Mesh (PPPM) scheme 
\cite{Hockney1981} which improves the force resolution by appending 
the gravitational force obtained with the Particle--Mesh (PM) scheme with a 
short-range Particle--Particle (PP) force. In CUBEP$^3$M code, the 
PM scheme is further split into two-level PM calculation to reduce 
the MPI communication required in solving the gravitational 
potential. In the TianNu simulation, CDM particles are initialized
at the cosmological redshift of 100, when the age of the Universe is 
16 million years. The neutrino particles are placed later when the system 
has evolved over 1 billion years. The two components are then evolved to 
the present Universe. Their CUBEP$^3$M code 
achieves 72\% weak scaling efficiency on 13,824 computational nodes 
(331,776 cores) of Tianhe-2 supercomputer, and the total wall clock 
time to complete their simulation is 52 hours.

\section{Innovations Realized}

\subsection{Vlasov Simulation in Six-Dimensional Phase Space and 
Combination with $N$-body Simulation}

\subsubsection{Vlasov Simulation}

One of our main innovations is completion of Vlasov simulation of
collisionless self-gravitating matter in the six-dimensional phase space. 
Historically, Vlasov simulations have been used in studies of collisionless plasma
as well as collisionless self-gravitating systems 
\cite{Janin1971, Cuperman1971, Fujiwara1981, Fujiwara1983b}. Unfortunately,
the applications are limited only to problems with spatially one- or
two-dimensions, because of an extremely large amount of memory and computational cost
necessary even for spatially two-dimensional problems. Five-dimensional gyrokinetic Vlasov 
simulations performed in numerical simulations of low-beta plasma 
\cite{Watanabe2005,Idomura2008} are the ones with the highest dimensionality
ever conducted. The advent of exaflop-class supercomputers, together 
with significant advances in numerical techniques, finally allows us 
to perform Vlasov simulations in the full six-dimensional phase space.

The Vlasov equation~(\ref{eq:vlasov}) is solved with a spatially
fifth-order finite volume method (see \S\ref{sec:SLMPP}). The
six-dimensional phase space volume is discretized on a uniform Cartesian
grid both in the spatial and velocity domains. The number of grids in
the spatial and velocity spaces are referred to as $N_{\rm x}$ and
$N_{\rm u}$, respectively. We adopt the directional splitting method
\cite{Cheng1976}, in which the Vlasov equation~(\ref{eq:vlasov}) is
split into six one-dimensional advection equations: three in the
physical space
\begin{equation}
 \label{eq:advec_pos}
 \pdif{f}{t} + \frac{u_i}{a(t)^2}\pdif{f}{x_i} = 0 \,\,\,\,\,(i=1,2,3)
\end{equation}
and another set of three advection equations in the velocity space
\begin{equation}
 \label{eq:advec_vel}
 \pdif{f}{t} - \pdif{\phi}{x_i}\pdif{f}{u_i} = 0 \,\,\,\,\,(i=1,2,3),
\end{equation}
where $(x_1, x_2, x_3)=(x,y,z)$ and $(u_1, u_2, u_3)=(u_x, u_y,
u_z)$. 

The time evolution of the distribution function from $t=t^n$ to
$t^{n+1}=t^n+\Delta t$ is performed as
\begin{equation}
 \begin{split}
 f(&\itbold{x}, \itbold{u},t^{n+1}) = D_{u_z}(\Delta t/2)D_{u_y}(\Delta t/2)D_{u_x}(\Delta t/2)  \\
  &\times D_x(\Delta t)D_y(\Delta t)D_z(\Delta t)\\
  &\times D_{u_z}(\Delta/2)D_{u_y}(\Delta t/2)D_{u_x}(\Delta t/2) f(\itbold{x}, \itbold{u}, t^n),
 \end{split}
\end{equation}
where $D_\ell(\Delta t)$ denotes an operator to advance an advection
equation along $\ell$-direction. Details of the numerical scheme to
advance an advection equation and its implementation are described in
\S\ref{sec:SLMPP} and \S\ref{sec:SIMD}, respectively. In our implementation, we adopt the single precision
floating point arithmetics for the Vlasov simulations.

\subsubsection{Combination with $N$-body Simulation}
Our simulations follow structure formation in a realistic,
observationally motivated cosmological model
where there exist both 
a dynamically cold component (CDM) and a hot thermal relic
(massive neutrino) that mutually interact through gravity.  
Therefore, the dynamics of CDM and massive neutrinos need to be solved 
simultaneously in a fully-coupled and self-consistent manner.  It is 
important to realize that the CDM component can be appropriately treated by a conventional
$N$-body method, because CDM is literally ``cold'' and has a very compact 
distribution in the velocity-space initially.  
We thus devise a hybrid of $N$-body and
Vlasov approaches, in which we adopt a sophisticated $N$-body method to 
solve the equation of motion of $N$-body particles that
represent the CDM component, whereas we directly integrate the Vlasov
equation (\ref{eq:vlasov}) for the massive neutrinos. Note that both of
the CDM and neutrino components share the common gravitational
potential; the mass density field $\rho(\itbold{x}, t)$ in
equation (\ref{eq:poisson}) is the sum of CDM and massive neutrinos.
The mass density of CDM is obtained from the distribution of $N$-body particles, 
and that of massive neutrinos is obtained by integrating the distribution 
function over the entire velocity space.

We employ the TreePM (Tree Particle-Mesh) method \cite{Bagla2002,Dubinski2004} to 
perform the $N$-body simulation for the CDM component. The TreePM scheme splits the gravitational 
force into two parts, short- and long-range forces each of which is computed with
the tree and particle-mesh (PM) schemes, respectively. In the PM scheme, the gravitational potential
is computed on a regular mesh grid (hereafter, the PM mesh grid) for  
the mass density field contributed by the CDM component {\it and} by 
the massive neutrinos. The long-range gravitational force at an arbitrary 
position is computed by differentiating and interpolating the gravitational 
potential defined on the PM mesh grid. Since we impose periodic boundary 
conditions, we solve the Poisson equation~(\ref{eq:poisson}) with the convolution
method \cite{Hockney1981} using Fast Fourier Transform (FFT). The short-range 
gravitational forces between $N$-body particles are 
computed with the tree algorithm to improve the force resolution in the high 
density regions which is otherwise missed in the conventional PM scheme. The 
calculation of the short-range forces is computed by a highly optimized gravity 
kernel, in which the force calculation is accelerated with the aid 
of SIMD instructions. It is originally developed for x86 architecture with 
SSE and AVX instruction sets (see \cite{Nitadori2006,Tanikawa2013}) and is named 
``Phantom-GRAPE'' after the API compatibility to GRAPE-5\cite{Kawai2000}. 
We port Phantom-GRAPE to Fujitsu A64FX processors on Fugaku supercomputer 
using the SIMD instruction set available on an A64FX processor, the Scalable Vector 
Extension (SVE) instruction set. The details of the implementation using SIMD 
instructions can be found in \cite{Tanikawa2013}. With the aid of the SVE 
instruction set, we achieve the performance of $1.2\times 10^9$ 
interactions/sec on a single core of a A64FX processor, whereas that of implementation 
without explicit use of the SVE instruction set is $2.4\times 10^7$ interactions/sec.  

For a simulation with $N_{\rm CDM}$ particles for the CDM component, we set the number of the PM mesh grid $N_{\rm PM}$ to $N_{\rm PM}=N_{\rm CDM}/3^3$ so 
that the elapsed time required for the $N$-body part is the shortest. 
We note that the positions and velocities of the
$N$-body particles are represented by double precision floating point
numbers.

\subsubsection{Domain Decomposition}
We consider a six-dimensional phase space
domain defined on $0 \le x,y,z \le L$
and $-V \le u_x, u_y, u_z < V$ in the Cartesian coordinate.  We evenly
decompose the physical space along each spatial axis for parallelization with MPI, but the velocity space is
not decomposed. Each spatial grid point holds an entire mesh grid for
the velocity space so that the calculation of the velocity moments of
the distribution functions such as mass density, mean velocity and
velocity dispersion tensor can be performed {\it without} any data transfer
among MPI processes. This efficient strategy helps us with improving the
overall performance of our code.
In what follows, let us denote the numbers of decomposed sub-domain as $n_x$,
$n_y$ and $n_z$ per side along $x$, $y$ and $z$ axes, respectively, and also the number of MPI processes as $N_{\rm proc}=n_x n_y n_z$.

In the $N$-body calculation part, the distribution of $N$-body particles
is decomposed into evenly spaced $n_x \times n_y \times n_z$ regions.
As for the PM scheme to compute the long-range gravitational forces, the CDM 
density field is computed on the PM mesh grids in each three-dimensionally 
decomposed domain, then is transferred among MPI processes so that the entire 
density field is decomposed into a two-dimensional manner, because
the efficient parallel three-dimensional FFT software library in the Fujitsu
SSL II/MPI package available on Fugaku supercomputer supports the two-dimensionally 
decomposed data layout. Aside from the parallel FFT, the MPI data communication
in $N$-body part mainly takes place in computing the mass density field 
contributed by the $N$-body particles and also in computing the short-range 
forces of the $N$-body particles with the tree method, both of which require
$N$-body particle distribution in the vicinity of adjacent domain boundaries.  

\subsection{Spatially High-Order Advection Scheme With A Single-Stage Time Integration\label{sec:SLMPP}}

One of the potential drawbacks of our Vlasov simulation is the large amount of
memory required to configure mesh grids not only in the physical space
but also in the velocity space. Thus, the spatial resolution of Vlasov
simulations is limited compared to conventional $N$-body simulations,
even with currently available state-of-the-art supercomputers. It is not
practical to improve the spatial and/or velocity resolutions by 
simply increasing the number of mesh grids.
Thus, it is important to adopt a numerical
scheme with spatially high-order accuracy and to effectively improve the
spatial resolution for a given number of mesh grid. It would be also ideal to
satisfy both monotonicity and positivity of numerical solutions
considering the physical and mathematical characteristics of the Vlasov
equation~(\ref{eq:vlasov}) and advection
equations~(\ref{eq:advec_pos},\ref{eq:advec_vel}). Note that numerical
advection schemes with a spatially high-order accuracy generally require
high-order temporal accuracy as well, in order to obtain numerically
stable solutions.  Hence one usually adopts a time integration scheme
with multiple stages such as temporally high order TVD Runge-Kutta
schemes \cite{Shu1988} at the expense of increased computational costs.

To realize spatially high order scheme with less computational cost,
we devise and adopt a novel numerical scheme, SL-MPP5 \cite{Tanaka2017}, 
which has spatially fifth-order accuracy with the monotonicity and 
positivity (MP) preservation and a temporally high-order time 
integration scheme with only a single stage. 
The coexistence of a spatially high order MP preserving schemes and a single stage
time integration scheme is realized for the first time in our new scheme by replacing
the polynomially reconstructed numerical fluxes at mesh boundaries in
the standard MP preserving scheme \cite{Suresh1997} with the ones constructed with the conservative
semi-Lagrange schemes \cite{Qiu2010, Qiu2011}. With this prescription,
we are able to obtain numerically stable solutions with spatially high order
accuracy using computationally less expensive time integration
scheme. This results in significant reduction of the overall computational
cost of the Vlasov simulation.  Spatially fifth order schemes 
with conventional time integration schemes usually require temporally third order
time integration schemes. In other words, it would be 
necessary to perform calculations of numerical fluxes three times 
per step. Our new scheme requires the calculation of numerical
fluxes only once per time step, and thus reduces the computational cost drastically.



\subsection{Efficient SIMD Vectorization in Vlasov Simulation\label{sec:SIMD}}
In order to realize the best possible performance on modern processor architectures,
SIMD vectorization is indispensable for optimization. Fujitsu A64FX processor in Fugaku supercomputer also has the SIMD instruction set named
Scalable Vector Extension (SVE), and can perform eight and 16 operations
of 64-bit and 32-bit data elements in parallel, respectively. We
explicitly utilize the SIMD instructions in implementing the advection
schemes described in \S\ref{sec:SLMPP}.

\begin{figure}[t]
 \centering
 \includegraphics[width=7.5cm]{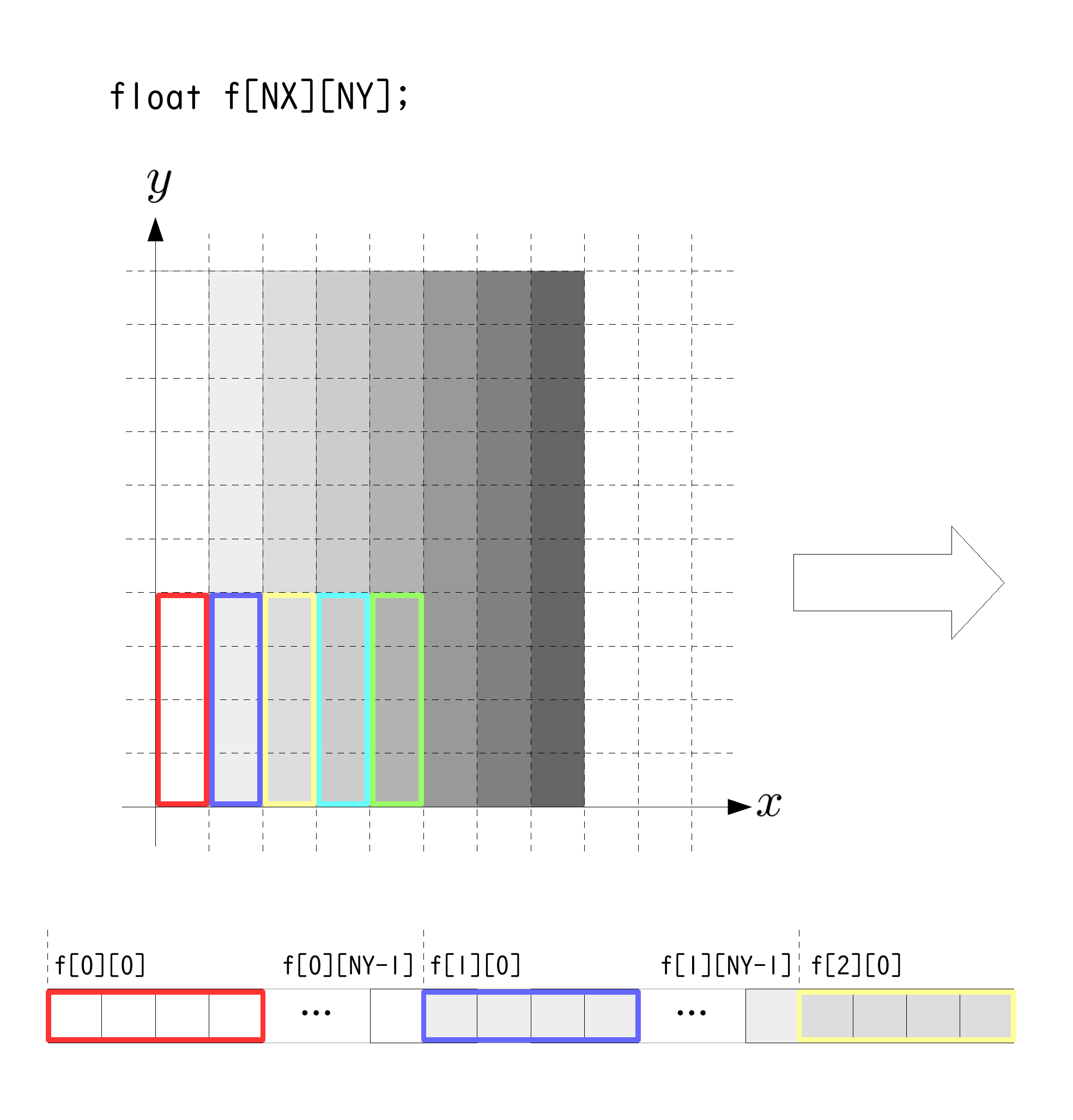}

 \caption{Schematic illustration of SIMD vectorization in advancing the
 Equation~(\ref{eq:advec_2d_x}). Colored boxes show the data layout
 loaded to individual SIMD registers, where the vector width is set to
 four. Note that the data in a SIMD register have continuous memory
 addresses. See the data layout in the bottom
 panel. \label{fig:advec_2d_x}}
\end{figure}

\begin{figure}[t]
 \centering \includegraphics[width=7.5cm]{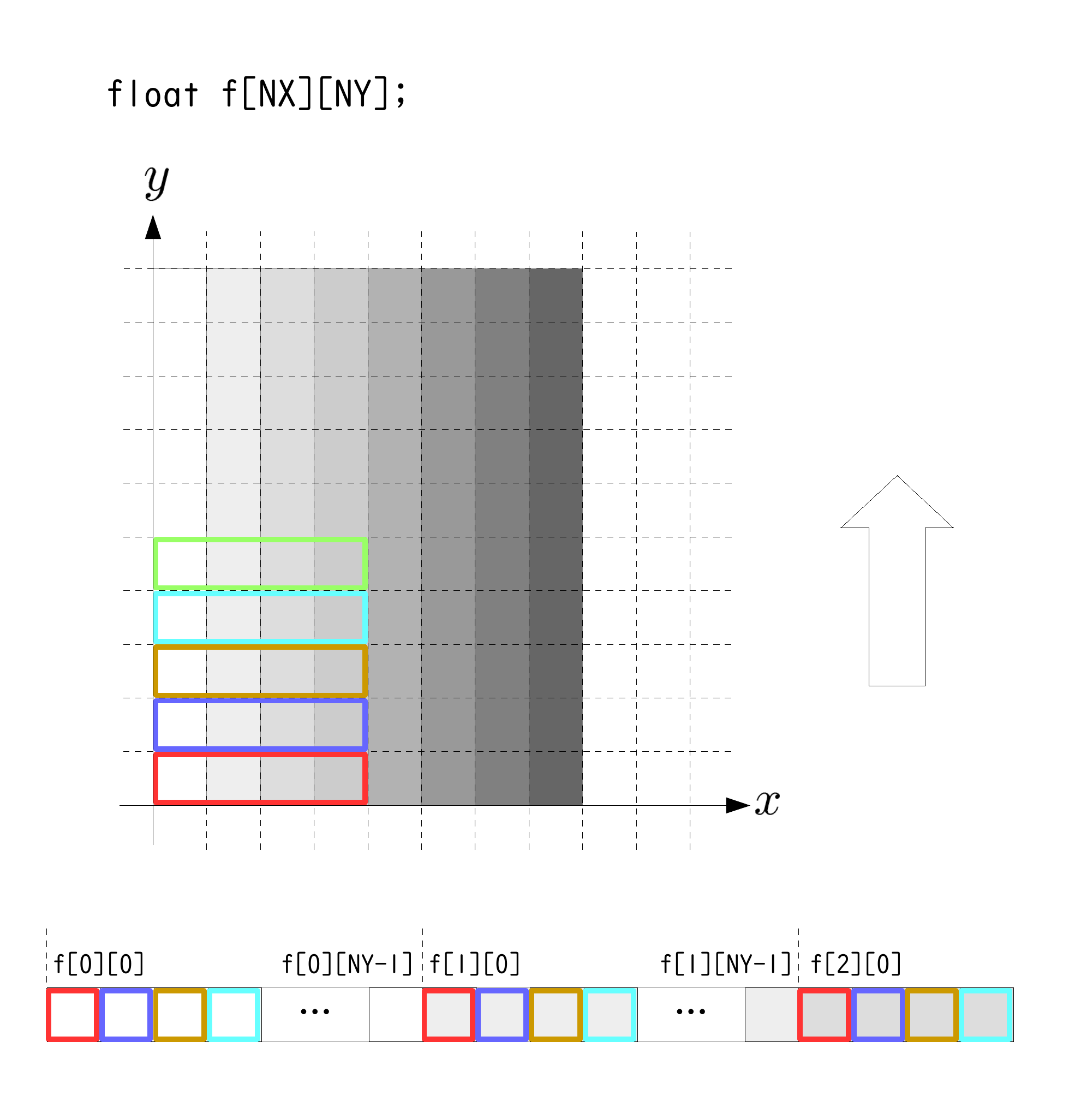}
 \caption{Illustration of the data layout on SIMD registers required to
 perform SIMD vectorization in advancing (\ref{eq:advec_2d_y}) along the
 $y$-axis for multiple columns. Data in discontinuous addresses should
 be packed into individual SIMD registers (colored
 boxes). \label{fig:advec_2d_y}}
\end{figure}

Here, we describe our approach in two dimensions as
a clear-cut case. It can be readily extended to six-dimensional
cases. Let us consider a two-dimensional advection equation
\begin{equation}
 \label{eq:advec_2d}
 \pdif{f(x,y,t)}{t} + v_x \pdif{f(x,y,t)}{x}+ v_y \pdif{f(x,y,t)}{y}=0,
\end{equation}
where $v_x$ and $v_y$ are the advection velocities along $x$ and $y$
axes, respectively. We adopt the directional splitting method to solve
this equation; we sequentially advance an advection equation along
$x$-direction
\begin{equation}
 \label{eq:advec_2d_x}
  \pdif{f(x,y,t)}{t} + v_x \pdif{f(x,y,t)}{x} = 0
\end{equation}
and one along $y$-direction
\begin{equation}
 \label{eq:advec_2d_y}
  \pdif{f(x,y,t)}{t} + v_y \pdif{f(x,y,t)}{y} = 0.
\end{equation}
Suppose that the function $f(x,y,t)$ is regularly discretized on the
$xy$-plane with the mesh grid as shown in
\figurename~\ref{fig:advec_2d_x}. In numerically advancing the advection
equation~(\ref{eq:advec_2d_x}) along the $x$-axis, it is straightforward
to perform the time integration for multiple rows with SIMD
instructions. Since the discretized data along the $y$-axis have
continuous memory addresses, the data aligned along the $y$-axis
(enclosed by each colored boxes in \figurename~\ref{fig:advec_2d_x}) can
be loaded to a SIMD register with a single instruction. We can then
solve (\ref{eq:advec_2d_x}) in parallel for multiple indices of the
$y$-coordinate with SIMD instructions.

The time integration along the $y$-axis with SIMD
instructions is not as simple as that along the $x$-axis. In order to
exploit SIMD instructions to integrate~(\ref{eq:advec_2d_y}) in parallel
for multiple columns, we need to load a set of data in discontinuous
memory addresses into SIMD registers as shown in
\figurename~\ref{fig:advec_2d_y}. This introduces significant overhead
of memory operations and hampers efficient SIMD workflows.

\begin{figure}[t]
 \centering \includegraphics[width=8.8cm]{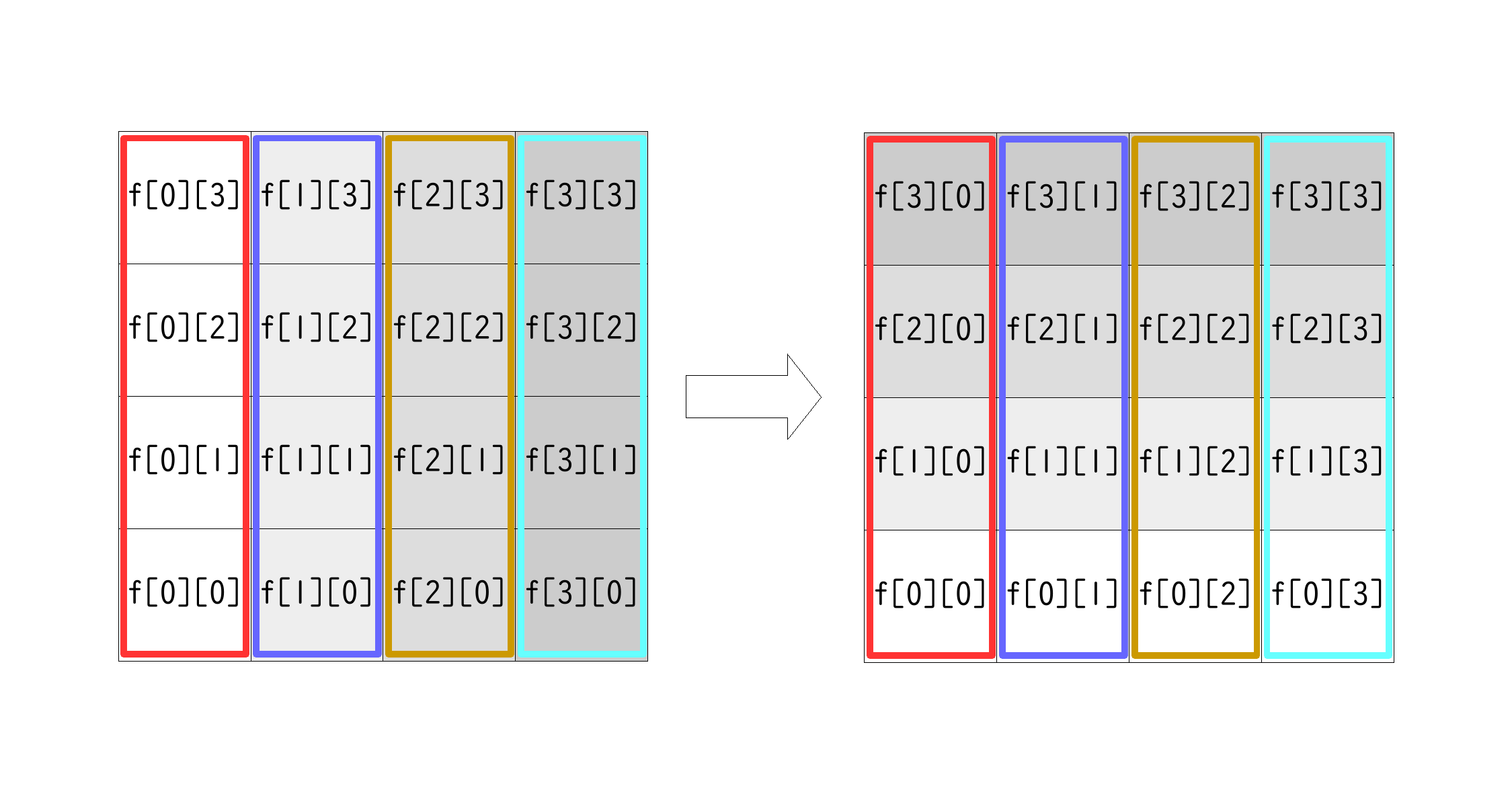}
 \caption{Transpose of 4$\times 4$ elements on four SIMD registers to re-arrange
 data layout. This is optimal to perform parallel advancing of equation
 (\ref{eq:advec_2d_y}) with SIMD instructions. Colored boxes indicate
 SIMD registers.\label{fig:transpose}}
\end{figure}

We utilize an efficient approach named ``load and transpose'' (LAT) method
to use the SIMD instruction set in solving
equation (\ref{eq:advec_2d_y}) along the $y$-axis.  First, we load the data
along the $y$-axis in the same manner as integrating (\ref{eq:advec_2d_x})
along the $x$-axis, as shown in the left panel of \figurename~\ref{fig:transpose}. 
In the case with the SIMD width of
$n$, the discretized data in $n$ contiguous columns are loaded to $n$
SIMD registers. Then, the layout of $n\times n$ data elements on the $n$
SIMD registers are transposed as shown in the right panel of
\figurename~\ref{fig:transpose}. The transpose of data on SIMD registers
can be done ``in-place'' by repeatedly shuffling the data elements
between SIMD registers. 64 SIMD instructions is required to transpose
16$\times$16 data layout on 16 SIMD registers. The resulting data
layouts on the SIMD registers are the same as depicted in
\figurename~\ref{fig:advec_2d_y}, which are suitable to perform advancing
equation (\ref{eq:advec_2d_y}) in parallel for multiple columns with 
SIMD instructions. Since the shuffle operations on SIMD 
registers can be performed very quickly
compared with memory operations on cache and on the main memory, we can perform
the time integration of equation (\ref{eq:advec_2d_y}) using SIMD instructions
with a significantly small overhead of memory operations. 
The LAT method is effective in solving advection equations not only in
the two-dimensional space but also in higher dimensional cases, and
can be extended to our Vlasov simulations in the
six-dimensional phase space.

\vspace{-5mm}
\begin{lstlisting}[caption={Structure of discretized distribution function}, label={list:df}]
 struct _df {
    float dens, ux_mean, uy_mean, uz_mean;
    float dfv[NUX][NUY][NUZ];
 };

 struct _df *df = (struct _df *) \\
    malloc(sizeof(struct _df)*NX*NY*NZ);
\end{lstlisting}

\smallskip

When we solve the Vlasov equation,
the discretized six-dimensional distribution function 
is defined as shown in List~\ref{list:df}, where
\verb|NX|, \verb|NY| and \verb|NZ| are the numbers of spatial mesh grids
along $x$, $y$ and $z$-directions, and
\verb|NUX|, \verb|NUY| and \verb|NUZ| are those of velocity mesh grids
along $u_x$, $u_y$ and $u_z$ directions, respectively. Time
integration of the Vlasov equation along a direction is implemented in the
form of a sextuple loop. The SIMD vectorization in solving the advection
equations along all the directions except for the $u_z$-axis can be done
in the same manner as depicted in \figurename~\ref{fig:advec_2d_x} by
running the second innermost loop over the index associated with the
$u_z$-axis. Advection along the $u_z$-axis corresponds to the case
shown in \figurename~\ref{fig:advec_2d_y}.

\begin{table}[!t]
\renewcommand{\arraystretch}{1.3} 
\caption{Performance of Vlasov simulation per CMG with and without SIMD 
instructions and the LAT method.
\label{tab:vlasov_simd_performance}}
\begin{tabular}{cccc}
\toprule
Direction & w/o SIMD inst. & w/ SIMD inst. & w/ LAT method\\
\midrule
$u_x$ & 4.84 [Gflops] & 176.7 [Gflops] & -- \\
$u_y$ & 7.14 [Gflops] & 233.3 [Gflops] & -- \\
$u_z$ & 7.44 [Gflops] & 17.9 [Gflops] & 224.2 [Gflops] \\
$x$ & 5.51 [Gflops] & 150.0 [Gflops] & -- \\
$y$ & 6.88 [Gflops] & 154.1 [Gflops] & -- \\
$z$ & 6.50 [Gflops] & 149.2 [Gflops] & -- \\
\bottomrule
\end{tabular}
\end{table}

We show the performance gain with the aid of the SIMD instruction set
and the LAT method on a A64FX processor in
Table~\ref{tab:vlasov_simd_performance}. There, we list the performance per
core memory group (CMG, see below for the details) of A64FX processor (see
\S\ref{sec:how_performance_measured}) measured in a set of Vlasov
simulations with $N_{\rm x}=32^3$ and $N_{\rm u}=64^3$ performed on two
nodes with and without the explicit use of SIMD instructions and the LAT
method. Clearly, the explicit use of SIMD instructions
improves the performances by a factor of $30$ in the
velocity space except for the one along the $u_z$-axis, and by a factor
of $18$ in the physical space. Note that the performance along the
$u_z$-axis is significantly lower even with the explicit use of the SIMD
instructions. This is owing to the inefficient load operations
to SIMD registers. With the use of the LAT method in solving the
advection equation along the $u_z$-axis, we have significantly improved the
efficiency of data load into SIMD registers. The resulting
performance is as good as those along the other axes in the velocity
space.

It is clearly seen that the performances in the velocity space (the
upper three items in \tablename~\ref{tab:vlasov_simd_performance}) is 
better than those in the physical space (lower
three items). This is because the operations in the advection in the
physical space include the data copy from/to the ghost mesh grid for the
MPI communication. Therefore, the performance of
the velocity space advection 
can be regarded as an ``uncontaminated''
sustained performance of our scheme on a single CMG, and achieves $12-15$\% of
the theoretical peak performance in a single precision arithmetics (1.54 Tflops/CMG). 

\begin{figure*}[htbp]
\centering
\includegraphics[width=5.85cm]{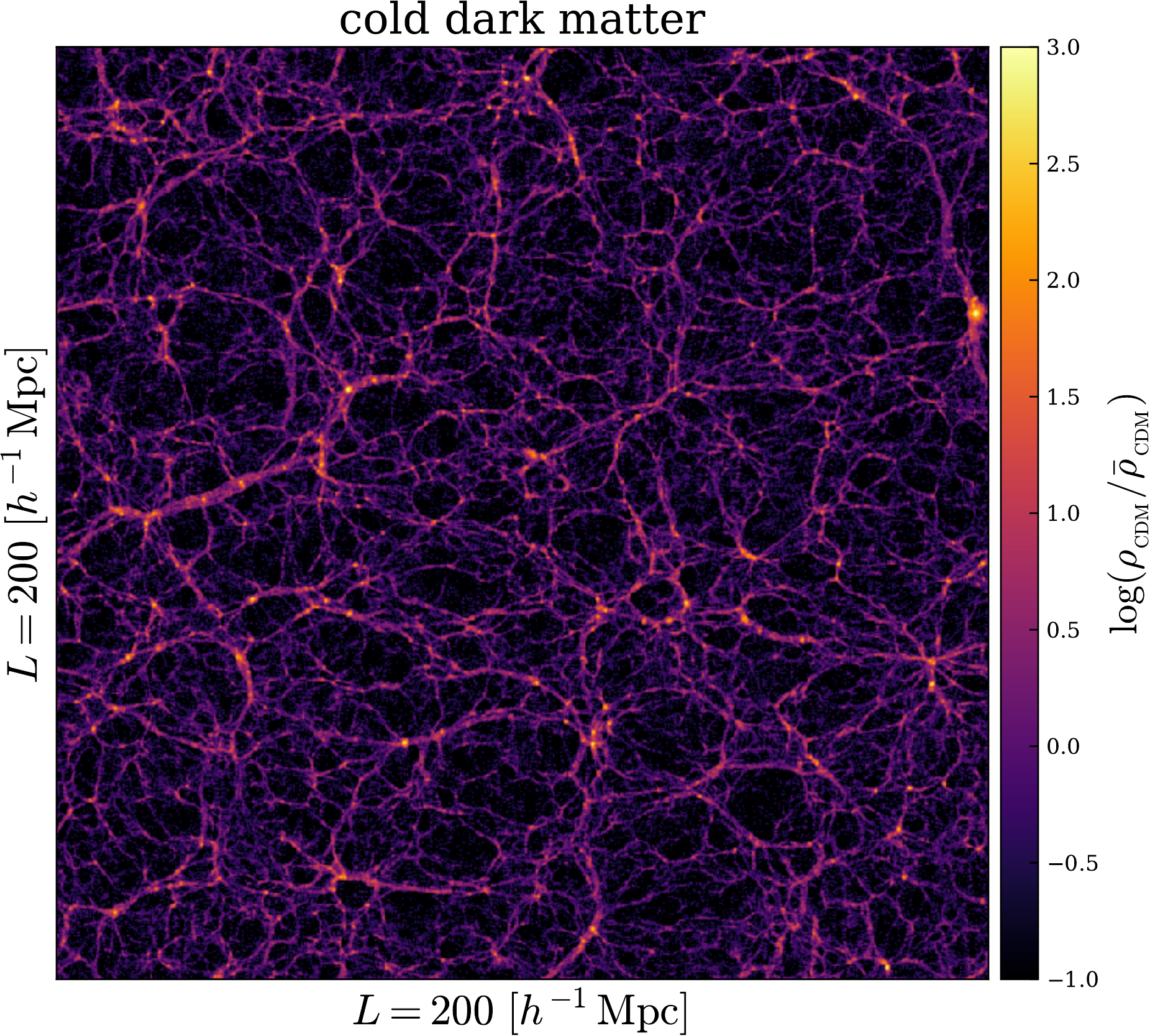}
\includegraphics[width=5.85cm]{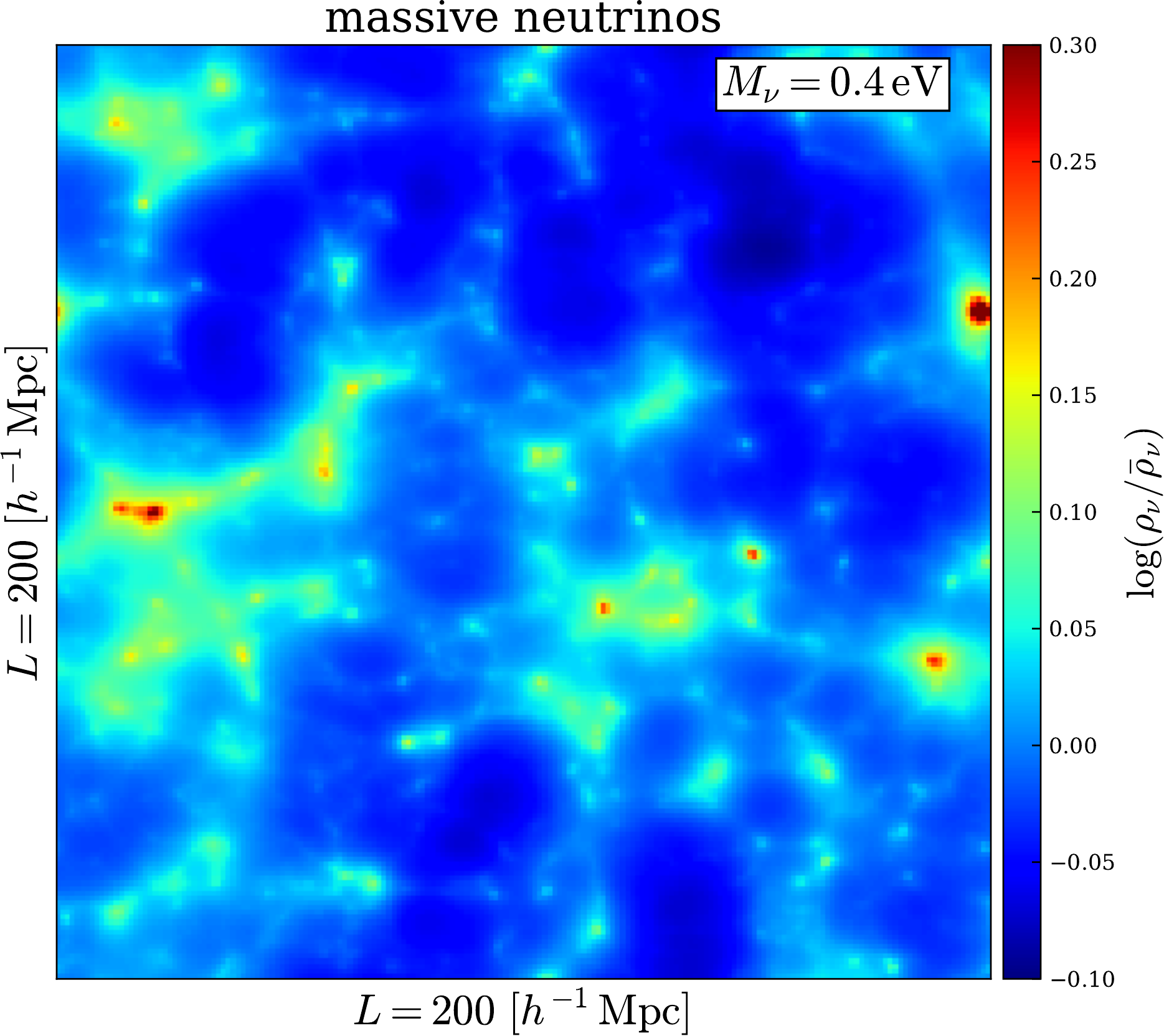}
\includegraphics[width=5.85cm]{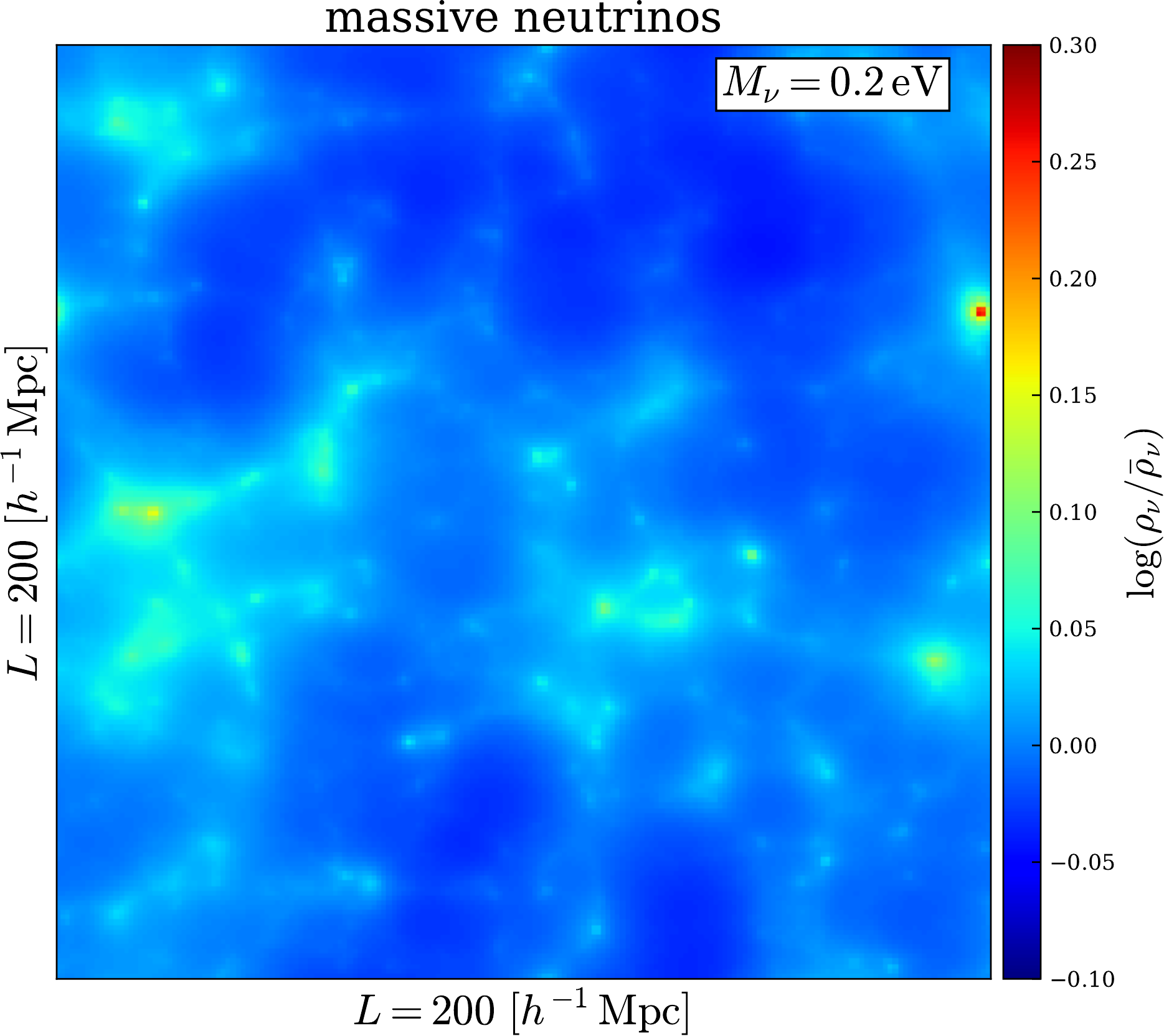}

\caption{Density maps of the CDM component and massive neutrinos simulated with Vlasov simulations. 
Our accurate Vlasov simulations are able to reproduce the difference in the large-scale distribution of 
massive neutrinos with mass of 0.4 eV (middle) and 0.2 eV (right).
\label{fig:nu_dens_map}}
\end{figure*}

\begin{figure}[t]
\centering

\includegraphics[width=8cm]{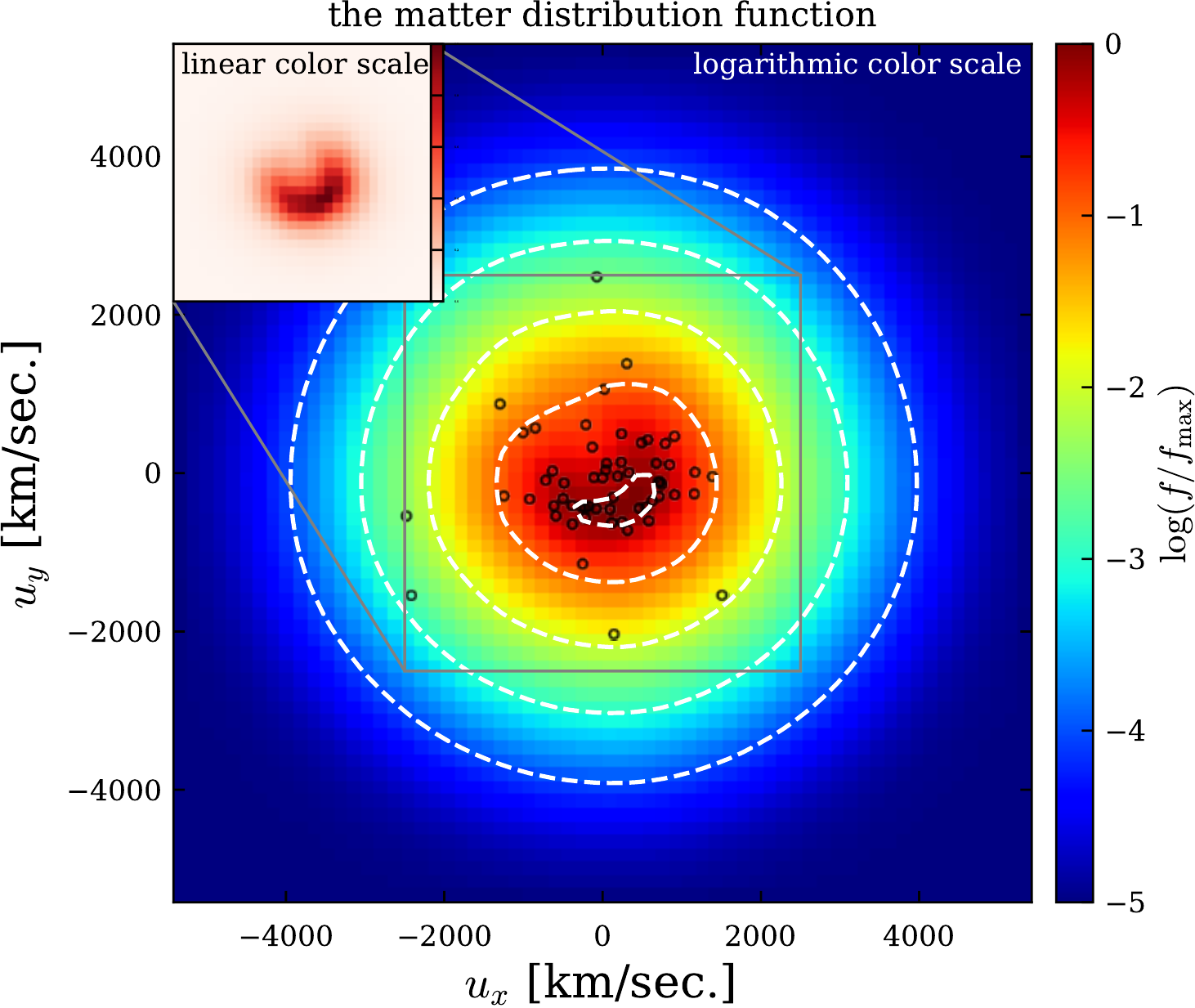}
\caption{The velocity distribution function of massive neutrinos at a 
single Vlasov mesh (physical position) in our Vlasov simulation (color). 
The inset shows the distribution in the low-velocity portion in linear 
color scale, showing deformed, fine structure in the velocity distribution.
Open circles are the neutrino particles in the same region in the corresponding 
particle-based simulation. \label{fig:vel_distribution}}
\end{figure}

\begin{figure*}[h]
\centering
\includegraphics[width=5.85cm]{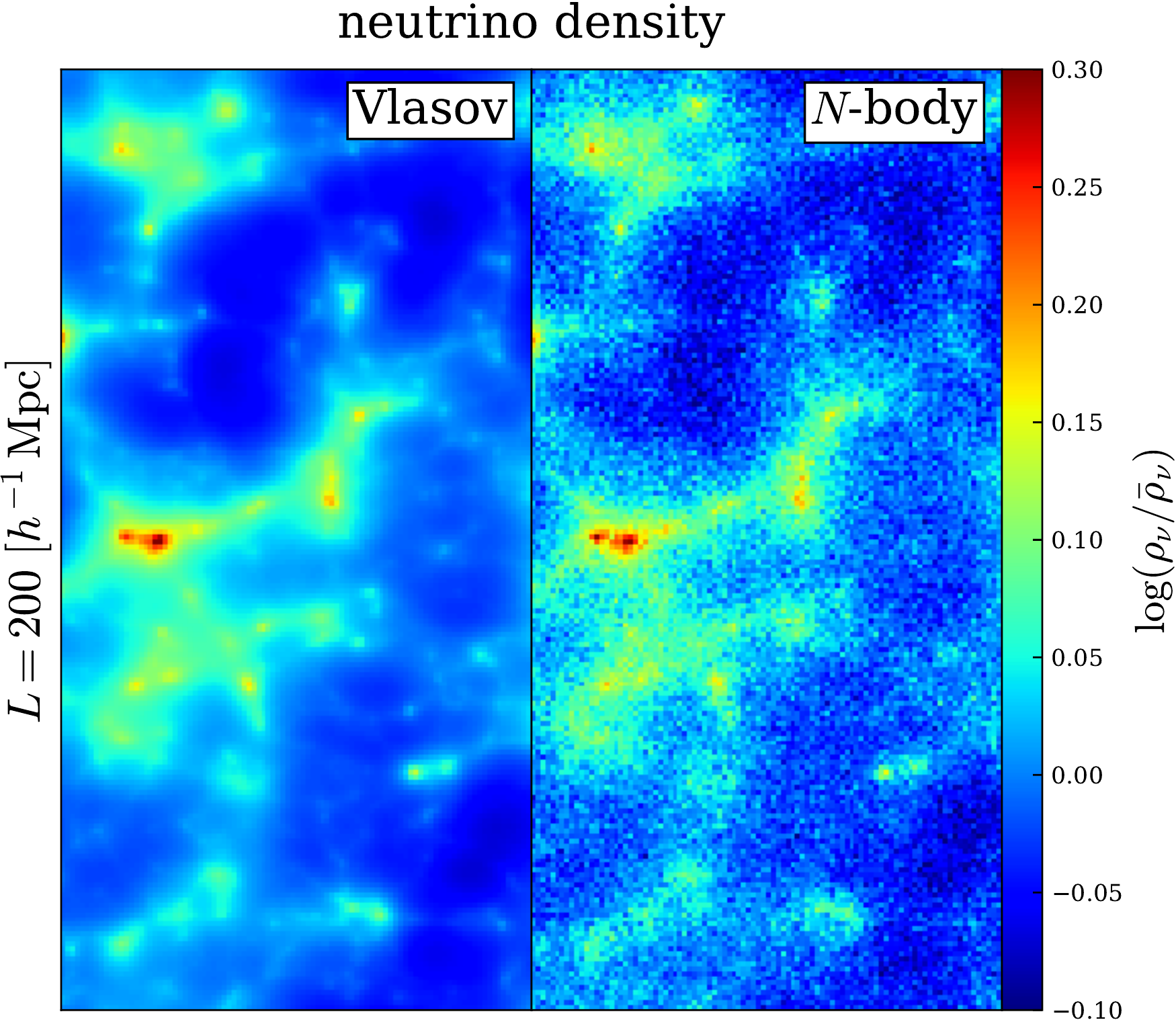}
\includegraphics[width=5.85cm]{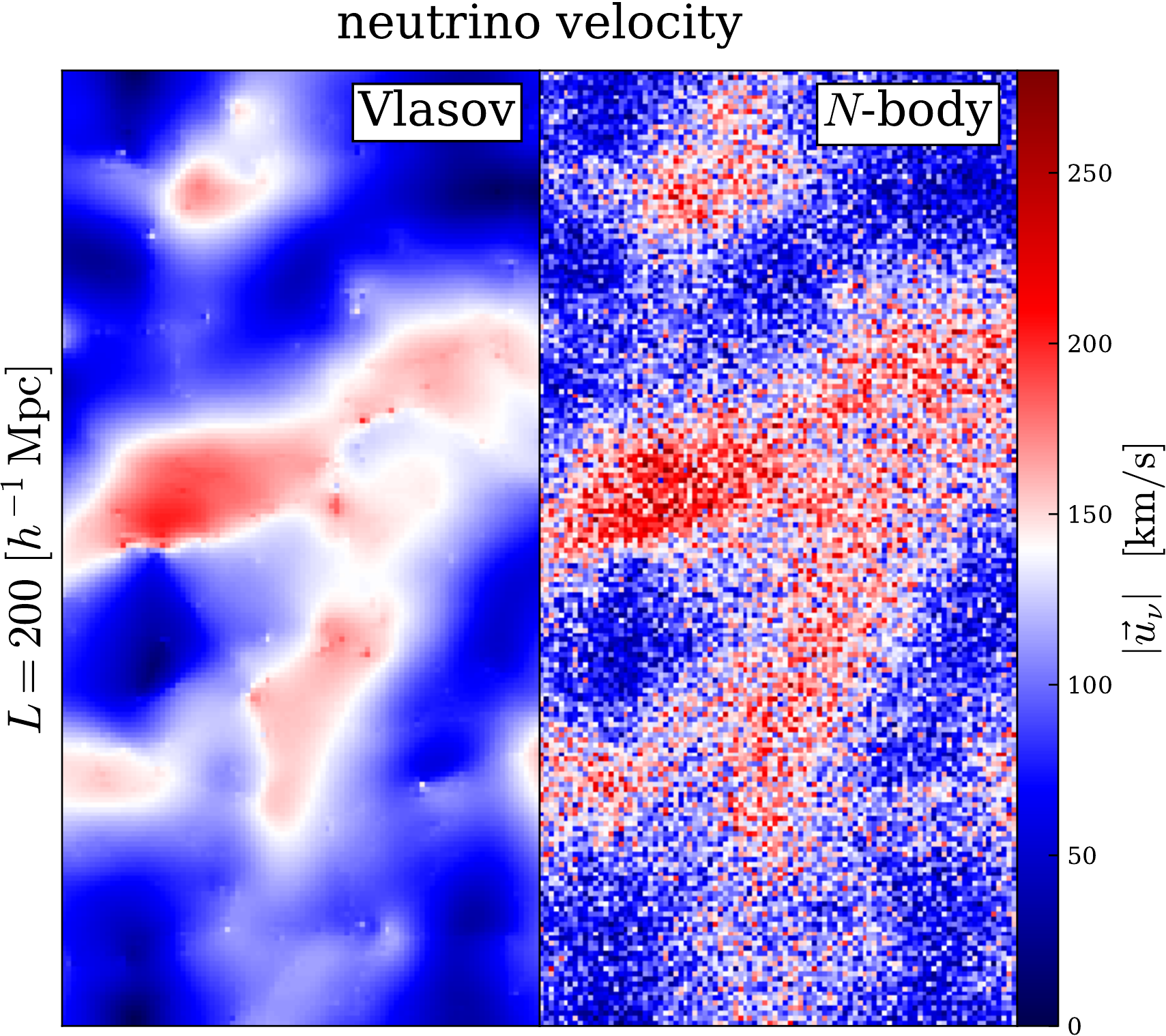}
\includegraphics[width=5.85cm]{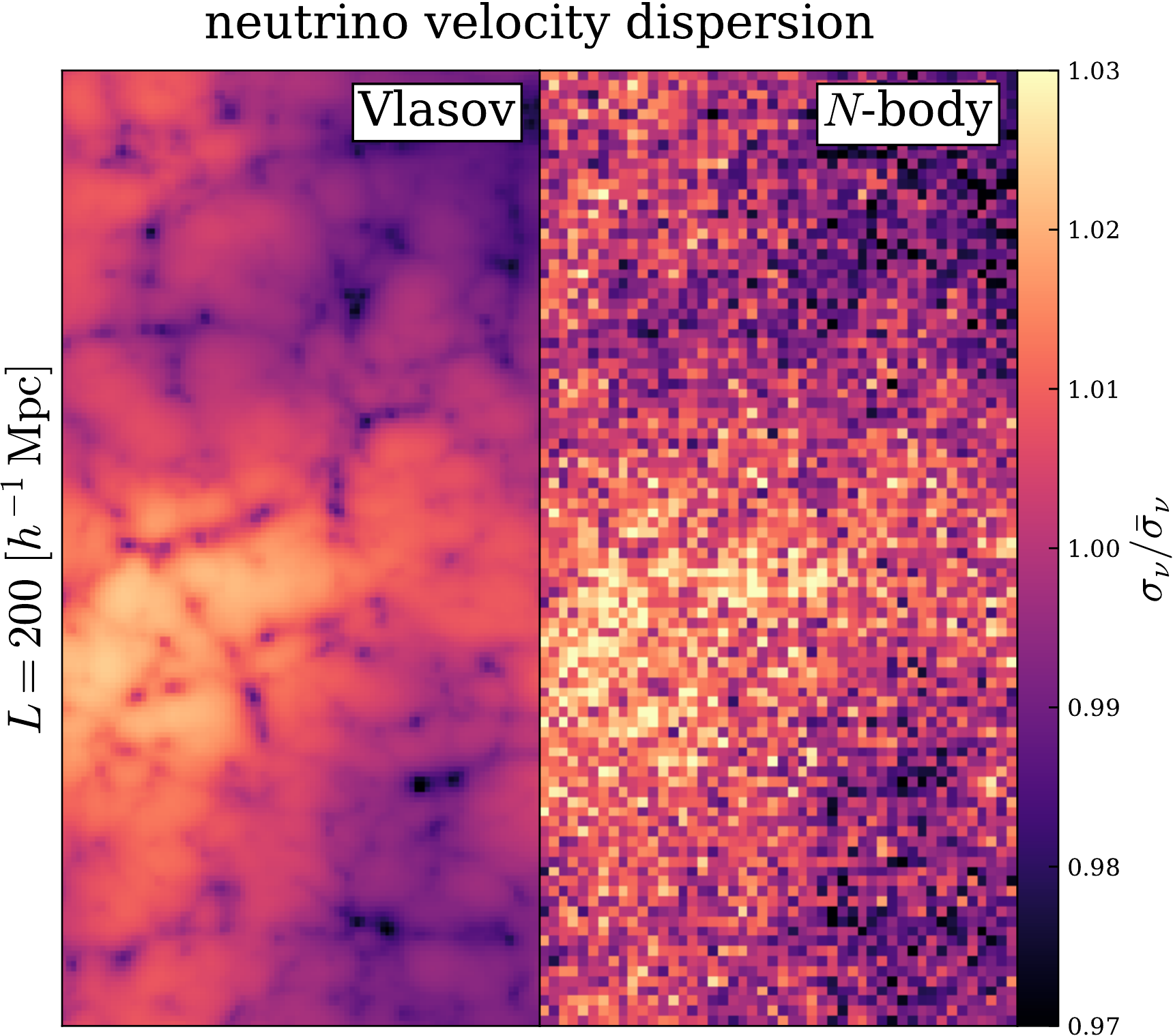}
\caption{Comparisons of mass density, velocity field and velocity dispersion of neutrinos between Vlasov and $N$-body simulations. Note that the map of velocity dispersion obtained in the $N$-body simulation is coarse-grained (smoothed) to reduce the shot-noise.\label{fig:neutrino_moment}}
\end{figure*}

\subsection{Superiority to $N$-body simulation\label{sec:result}}
\figurename~\ref{fig:nu_dens_map} compares the density field of the CDM component, massive 
neutrinos obtained with our hybrid Vlasov/$N$-body simulation (the run M24 listed
in \tablename~\ref{tab:scaling_runs}). 
The distribution of massive neutrinos is 
quite diffuse compared with that of the CDM component owing to their very large 
velocity dispersion. The neutrino distribution roughly traces that of CDM on a large scale, 
suggesting higher neutrino densities in and around high density regions of CDM. 
The smoother distribution of the neutrinos prevents the nonlinear growth of the
small-scale clustering  of CDM (and hence galaxies), which is expected to be 
observed by future galaxy surveys. We also show the density fields of massive neutrinos simulated with
different neutrino masses of $M_\nu = 0.2\,{\rm eV}$ and $0.4\,{\rm eV}$, where $M_\nu$ is the
sum over three mass eigenvalues of neutrinos. The distribution of massive neutrinos depends
on their mass $M_\nu$, and those of CDM and galaxies are strongly affected by the neutrino distribution.

\figurename~\ref{fig:vel_distribution} shows the local velocity distribution function of massive
neutrinos at a random position in our Vlasov/$N$-body simulation and the corresponding 
one in the $N$-body simulation starting from the equivalent initial condition. Our 
Vlasov/$N$-body simulation reproduces a smooth, long-tailed distribution as well as the deformation (substructure) in the 
low velocity patch, but the coarse sampling in the $N$-body simulation (denoted by
open circles) does not allow us even to discern such features. 

In \figurename~\ref{fig:neutrino_moment}, we show the comparison of density fields, velocity fields and velocity dispersion of massive neutrinos
simulated with one of our Vlasov/$N$-body hybrid simulations (the same as shown in \figurename~\ref{fig:nu_dens_map}) and
their counterparts obtained by an $N$-body simulation originated from the same initial condition, 
in which we employ $768^3$ particles for the CDM component and $8\times 768^3$ particles for the massive neutrinos.
The neutrino density field obtained with our Vlasov/$N$-body simulation is smooth and resolves fine structures
uniformly across the entire computational domain, whereas the counterpart 
in the $N$-body simulation is compromised by the shot noise; the fine structures resolved
in the Vlasov/$N$-body simulation are missed and heavily contaminated by the shot noise.
The poor representation of the velocity structure in the $N$-body simulation 
seen in \figurename~\ref{fig:vel_distribution} also affects 
higher order velocity moments of the distribution function, such as the velocity field and velocity 
dispersion more seriously, as can be seen in \figurename~\ref{fig:neutrino_moment}.
It should be noted that required wall time to complete these two simulations using the same amount of computational resources are 
almost comparable, indicating that our Vlasov/$N$-body simulation is clearly superior to 
conventional particle-based $N$-body ones in simulating the dynamics of massive neutrinos.

\section{How Performance Was Measured\label{sec:how_performance_measured}}

\subsection{Platform and Setup}

We perform our numerical simulations on Fugaku supercomputer
which consists of 158,976 computational nodes, 
each of which has an Fujitsu A64FX processor based on ARMv8-A ISA.
The A64FX processor has four sets of CMGs, each of which comprises of 
12 compute cores and 8 GB HBM2 memory, and thus 48 compute cores and 
32 GB memory in total. The four CMGs in a chip are connected via a 
ring bus network with a bandwidth of 115 GB/s. Theoretical peak 
performance per CMG is 0.77 and 1.54 Tflops for double and single 
precision arithmetics, respectively. Computational nodes are connected
via Tofu interconnect D, a six-dimensional torus network with
a mesh size of $24 \times 23 \times 24 \times 2 \times 3 \times 2$.  
In what follows, each MPI process is assigned to a single or two CMGs 
depending on the problem size. Therefore, 
the number of MPI processes is two or four times as many as the number of computational
nodes.  MPI processes are allocated on the six-dimensional torus 
network so that MPI communications between physically adjacent domains 
are kept fenced within a single hop. 

For the measurement of the scalability, we conduct numerical simulations with
the box size of $L=200 h^{-1}$ mega parsec (Mpc) per side for the standard 
cosmological model determined by the recent observation of the cosmic microwave
background (CMB) \cite{Planck2015XIII}. Here, $h$ is the normalized Hubble 
constant in units of 100 km/sec/Mpc. We assume the total mass of neutrino 
over three mass eigenstates to be 0.4 eV, which is close to the upper limit 
placed by the CMB observation \cite{Planck2015XIII}. 
The performance is evaluated in terms of wall clock elapsed time 
measured with the \verb|clock_gettime()| system call. For each run 
listed in \tablename~\ref{tab:scaling_runs}, we run the simulations by 40 steps 
and take the median values of the 40 measured elapsed times.  

As for the measurement of time-to-solution, we setup an initial condition
with the box size of 1200$h^{-1}$ Mpc at a cosmological redshift of 10, 
similar to that of the existing state-of-the-art simulation \cite{Emberson2017}. 
We measure the total end-to-end elapsed time including that for I/O with the 
\verb|clock_gettime()| system call.

\section{Performance Results}

In this section, we present the performance of our hybrid Vlasov/$N$-body
simulation in terms of scalability and time-to-solution. 
Table~\ref{tab:scaling_runs} lists the parameters of runs presented 
in this section, where we show the number of mesh grids in Vlasov 
simulation and $N$-body particles, the number of computational nodes, 
the number of MPI processes along each axis of domain decomposition, 
and the number of MPI processes per node. We adopt a naming convention of these runs
in which the first letters S, M, L, H and U
denote the number of spatial mesh grids of the Vlasov simulation 
$N_{\rm x}=96^3$, $192^3$, $384^3$, $768^3$ and $1152^3$, respectively,
followed by the number of computational nodes in units of 144 nodes. 
The number of $N$-body particles for the CDM component is proportional
to that of Vlasov mesh grids as $N_{\rm CDM}=9^3N_{\rm x}$, except for
that of the largest run (U1024), in which $N_{\rm CDM}$ is same as 
that in the H run group and set to $N_{\rm CDM}=6912^3$. Note that H1024
and U1024 employ 147,456 computational nodes out of Fugaku's 
entire system (158,976 computational nodes), and thus they can be effectively 
regarded as full system runs of Fugaku supercomputer.

\begin{table}[htbp]
\renewcommand{\arraystretch}{1.4}
\caption{Runs for measurements of weak and strong scalings and
time-to-solution.\label{tab:scaling_runs}}
\begin{tabular}{llcccc}
\toprule
\rule[-4mm]{0pt}{9mm} ID &$(N_{\rm x}$, $N_{\rm u})$ & $N_{\rm CDM}$ & $N_{\rm node}$ & $(n_x, n_y, n_z)$ & $\displaystyle \frac{N_{\rm proc}}{N_{\rm node}}$\\
\midrule
S1 & $(96^3, 64^3)$ & $864^3$ & 144 & $(12, 12, 2)$ & 2\\

S2 & $(96^3, 64^3)$ & $864^3$ & 288 & $(12, 12, 4)$  & 2\\

S4 & $(96^3, 64^3)$ & $864^3$ & 576 & $(12, 12, 8)$ &2 \\

M8 & $(192^3, 64^3)$ & $1728^3$ & 1152 & $(24, 24, 4)$ & 2\\

M12 & $(192^3, 64^3)$ & $1728^3$ & 1728 & $(24, 24, 6)$ & 2 \\

M16 & $(192^3, 64^3)$ & $1728^3$ & 2304 & $(24, 24, 8)$ & 2\\

M24 & $(192^3, 64^3)$ & $1728^3$ & 3456 & $(24, 24, 12)$ & 2\\

M32 & $(192^3, 64^3)$ & $1728^3$ & 3456 & $(24, 24, 16)$ & 2\\

L48 & $(384^3, 64^3)$ & $3456^3$ & 6912 & $(48, 48, 6)$ & 2\\

L64 & $(384^3, 64^3)$ & $3456^3$ & 9216 & $(48, 48, 8)$ & 2\\

L96 & $(384^3, 64^3)$ & $3456^3$ & 13824 & $(48, 48, 12)$ & 2\\

L128 & $(384^3, 64^3)$ & $3456^3$ & 18432 & $(48, 48, 16)$ & 2\\

L256 & $(384^3, 64^3)$ & $3456^3$ & 36864 & $(48, 48, 32)$ & 2\\

H384 & $(768^3, 64^3)$ & $6912^3$ & 55296 & $(96,96,24)$ & 4\\

H512 & $(768^3, 64^3)$ & $6912^3$ & 73728 & $(96,96,32)$ & 4\\

H768 & $(768^3, 64^3)$ & $6912^3$ & 110592 & $(96,96,48)$ & 4\\

H1024 & $(768^3, 64^3)$ & $6912^3$ & 147456 & $(96,96,64)$ & 4\\

U1024 & $(1152^3, 64^3)$ & $6912^3$ & 147456 & $(48,48,128)$ & 2\\
\bottomrule
\end{tabular}
\end{table}

\subsection{Scalability}

\begin{figure*}[!t]
 \centering
 \includegraphics[width=14.5cm]{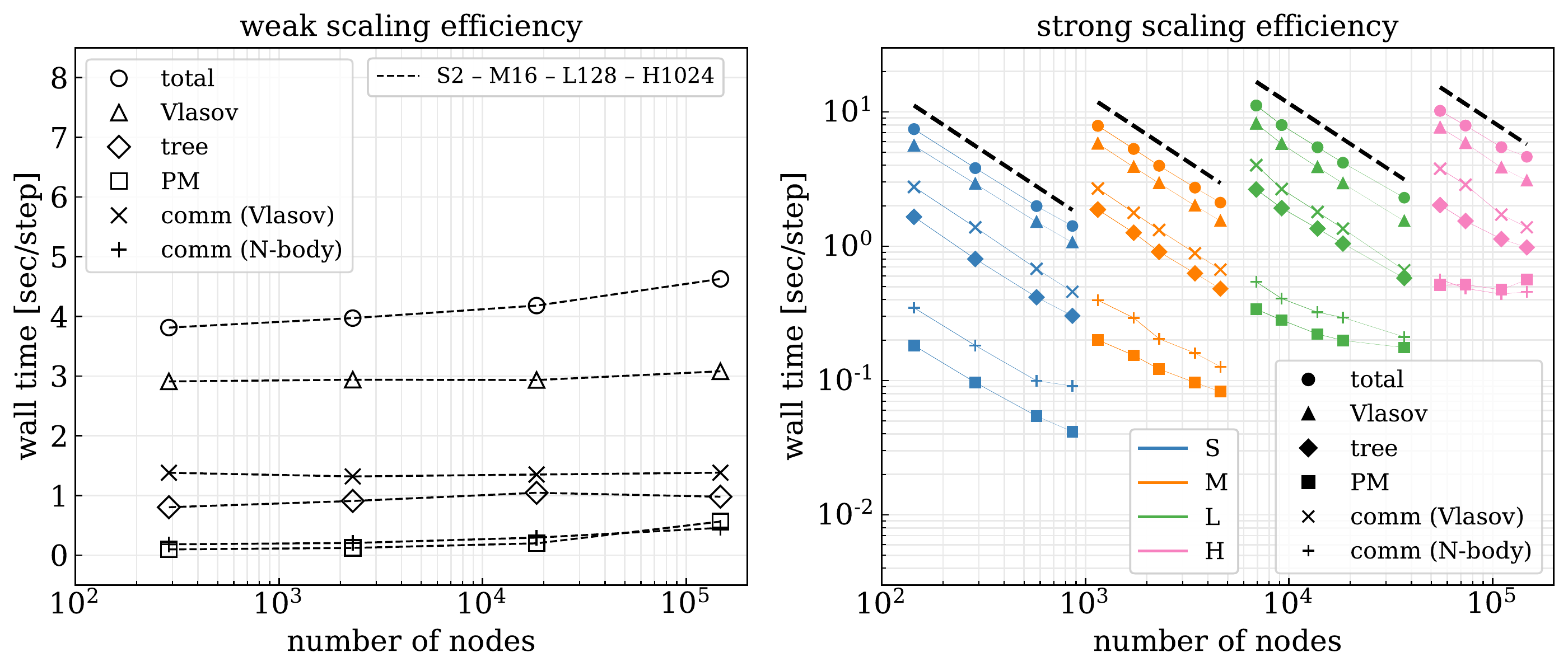}

 \caption{Weak (left) and strong (right) scaling efficiencies of Vlasov, 
 tree and PM parts as well as the total scaling efficiency. Dashed line 
 shows the ideal scaling. \label{fig:scaling}}
\end{figure*}

In order to measure the weak and strong scalings of our hybrid Vlasov/N-body 
simulations, we perform 17 runs in S, M, L and H run groups
listed in Table~\ref{tab:scaling_runs}.
We measure the elapsed time per step for integrating the Vlasov equation (Vlasov part),
for computing short-range forces of $N$-body particles using the tree method 
(tree part), and for solving the Poisson equation with the PM scheme (PM part), 
separately as well as the ones for communicating data between MPI processes 
required in the Vlasov and tree parts in a manner described in \S\ref{sec:how_performance_measured}. \figurename~\ref{fig:scaling} 
depicts the decomposed elapsed time per step as well as the total elapsed 
time per step measured against number of nodes for S, M, L and H run groups listed in 
\tablename~\ref{tab:scaling_runs}. The elapsed time for the Vlasov part amounts to about 70\% 
of the total, and is the most dominant in the whole simulation. 
In the left panel, we present the elapsed time of each part 
and that of the whole 
simulation for a sequences of runs, S2, M16, L128, and H1024. It shows 
a measure of weak scaling efficiency and summarized in \tablename~\ref{tab:weak_scaling}.
The weak scaling efficiency of the Vlasov part is higher than 90\% for up to nearly
full system (147,456 nodes) of Fugaku supercomputer. We note that the
scaling of the PM part is 
not excellent because the FFT calculations involved in the PM part is parallelized only in a two-dimensional manner with $n_x n_y$ MPI 
processes, although it has a minor impact on the whole performance. 

Comparisons of the elapsed time per step between runs in each of S, M, L and H 
run groups depicted in the right panel of \figurename~\ref{fig:scaling} 
show the strong scaling efficiencies for the Vlasov, tree and PM parts and the 
whole simulation. which are summarized in \tablename~\ref{tab:strong_scaling}.
The strong scaling efficiencies of
the most time-consuming Vlasov part are excellent and better than 90\% for 
M, L and H run groups. The PM part appears slightly less efficient, but it can be ascribed to the compromised 
parallel efficiency of the FFT calculation mentioned above. Note that 
the degree of parallelism for the FFT calculation, $n_x n_y$, is constant within each 
run group.
Despite this, the overall strong scaling efficiencies are excellent in 
all the run groups.

\begin{table}[t]
\renewcommand{\arraystretch}{1.3}
\caption{Weak scaling efficiencies for the whole and each part of the simulation.\label{tab:weak_scaling}}
\begin{tabular}{cccc}
\toprule
 & S2--M16 & S2--L128 & S2--H1024 \\
 \midrule
 total & 96.0 \% & 91.1\% & 82.3\% \\

 Vlasov & 99.0\% & 99.2\% & 94.4\% \\

 tree & 88.4\% & 76.8\% & 82.0\% \\

 PM & 79.5\% & 48.7\% & 17.1\% \\
 \bottomrule
\end{tabular}
\end{table}

\begin{table}[t]
\renewcommand{\arraystretch}{1.3}
\caption{Strong scaling efficiencies for the whole and each part of the simulation\label{tab:strong_scaling}}
\begin{tabular}{ccccc}
\toprule
& S & M &  L &  H \\
\midrule
total & 87.7\%& 93.3\% & 91.1\% & 82.4\%\\

Vlasov & 87.5\% & 93.9\% & 99.6\% & 93.0\%\\

tree & 90.9\% & 97.1\% & 85.7\% & 77.5\% \\

PM & 72.9\% & 60.6\% & 36.2\% & 34.1\% \\
\bottomrule
\end{tabular}
\end{table}

\subsection{Time-To-Solution}

No other simulations ever conducted can be directly compared with ours since this is
the first Vlasov simulation in the full six-dimensional phase space domain. 
As a time-to-solution reference, we choose the largest particle-based $N$-body 
simulation of massive neutrinos in the LSS formation, which
shares the common scientific motivation and numerical outcomes with our simulations.
The largest particle-based $N$-body simulation of the LSS formation with 
massive neutrinos was the TianNu simulation performed on Tianhe-2 supercomputer\cite{Emberson2017}. 
A wall clock time of 52 hours was required to complete the simulation with $6912^3$
CDM particles and $8\times 6912^3$ neutrino particles \cite{Emberson2017}.

It is not straightforward to compare the numerical results obtained from 
a particle-based $N$-body simulation and from our Vlasov simulation. It would be appropriate and fair
to examine the following two important quantities: the spatial resolution and the level 
of shot noise. In $N$-body simulations, important physical quantities such as density 
and velocity fields are calculated by averaging the mass and velocity 
of individual $N$-body particles over a certain volume or a certain number 
of particles. Smoothing over a large number of particles lowers the level 
of shot noise in the local physical quantities, but it inevitably compromises the effective spatial resolution. Simply, by averaging over $N_{\rm s}$ 
particles, one obtain the spatial resolution of 
$\Delta L \simeq N_{\rm s}^{1/3}\times L/N_{\nu}^{1/3}$,
where $L$ is the size of a cubic simulation box and $N_{\nu}$ is the number 
of particles for massive neutrinos, and the shot noise level is estimated to be $1/N_{\rm s}^{1/2}$. 
In terms of the signal-to-noise ratio $S/N$, it is related as $S/N=N_{\rm s}^{1/2}$
following the simple Poisson statistics. 
Thus, the largest TianNu $N$-body simulation has an effective spatial resolution of 
neutrino distribution given by
\begin{align}
\Delta L & = \frac{L}{13824}(S/N)^{2/3} \\ 
    & \simeq \frac{L}{640}\left(\frac{S/N}{100}\right)^{2/3}\simeq \frac{L}{1018}\left(\frac{S/N}{50}\right)^{2/3}
\end{align}
as a function of $S/N$. For a sufficiently small shot noise level
of 1\%, or equivalently $S/N=100$, for example, the effective spatial resolution is 
$\Delta L\simeq L/640$ and is almost the same as the resolution of our H run group 
with $N_{\rm x} = 768^3$. Hence the TianNu simulation can be regarded
to be ``equivalent'' to the H run group in terms of spatial resolution. 
If we conservatively
allow the shot noise level to be up to 2\% ($S/N=50$), the effective spatial resolution is 
$\Delta L = L/1018$, which corresponds to that of the U run group with $N_{\rm x} = 1152^3$.

We perform two end-to-end runs, H1024 and U1024, with $N_{\rm x}=768^3$ and $1152^3$, 
respectively, on 147,456 nodes, nearly full system of Fugaku supercomputer. 
The initial condition is set up at a redshift of $z=10$
with a size of simulation box of $1200h^{-1}$ Mpc, and evolved to the current 
Universe ($z=0$). It should be noted that the TianNu simulation introduces the 
dynamical effect of massive neutrinos after a redshift of $z=5$, later than
the epoch in our simulation. 
Also the superior resolution in the velocity space of our Vlasov simulation (Fig. 5 and 6) is not considered here.
Hence our simulation is more elaborate and accurate.
The end-to-end elapsed time to complete these simulations including I/O are 1.92 hours
(6183 seconds for the execution and 733 seconds for I/O) for the H1024 run and
5.86 hours (20342 seconds for the execution and 782 seconds for I/O) 
for the U1024 run, which are improved by a factor of 27 and 8.9, respectively,
making a great leap compared with the state-of-the-art TianNu $N$-body simulation.

\begin{figure*}[h]
\centering
\includegraphics[width=17.0cm]{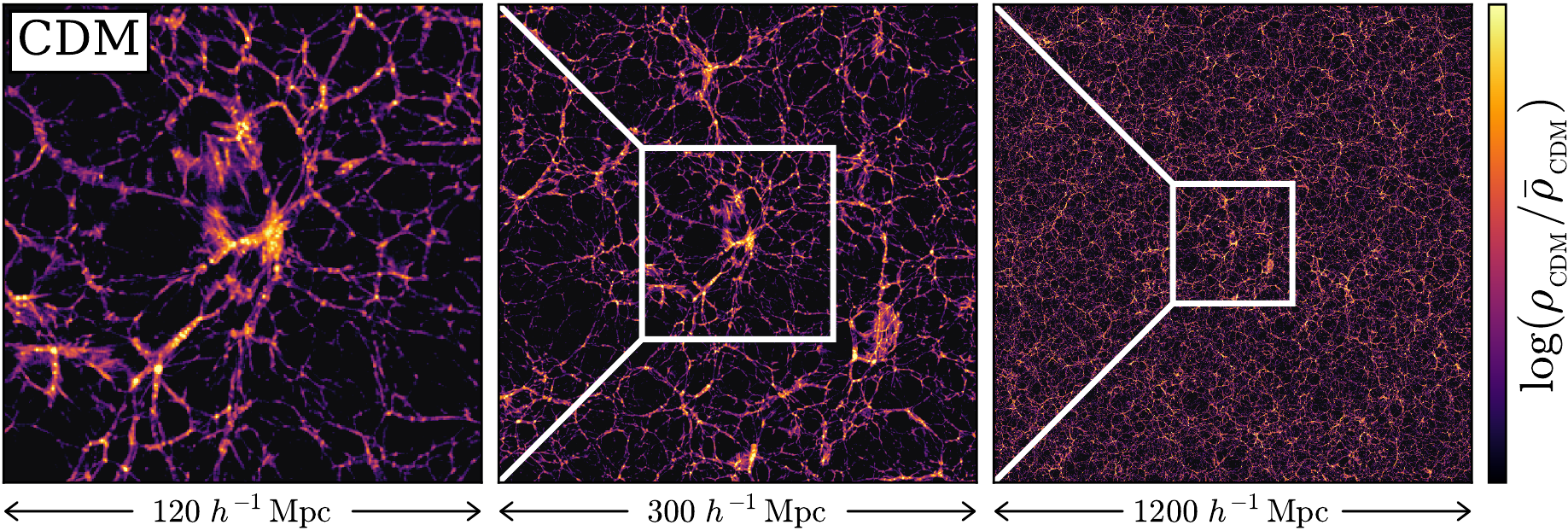}
\includegraphics[width=17.0cm]{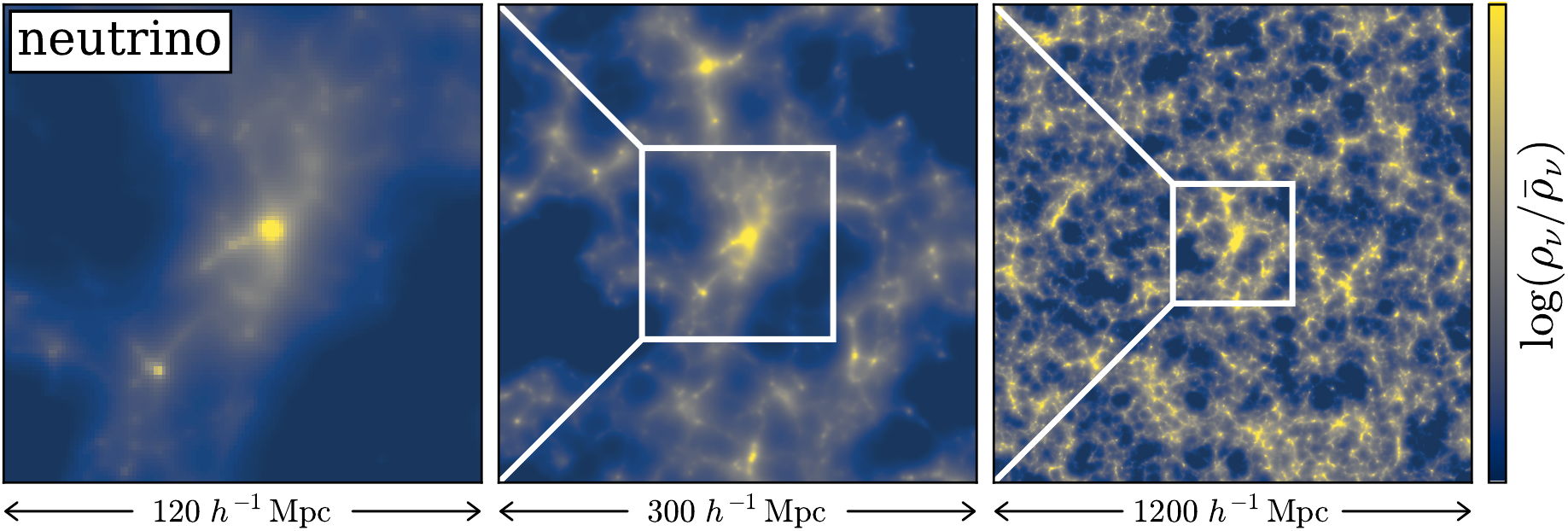}
\caption{Density maps of CDM (upper), and massive neutrinos (bottom) obtained 
in our largest Vlasov/$N$-body simulation (run ID U1024). \label{fig:H1024_dens}}
\end{figure*}

\section{Implications}

We have presented the results of the world's first and largest Vlasov 
simulation of massive neutrinos in the six-dimensional phase space coupled
with particle-based $N$-body simulation of 
cold dark matter in the context of cosmic
structure formation. Our simulation follows the gravitational dynamics of 
massive neutrinos in a self-consistent, fully coupled manner with the LSS 
formation. Our novel method provides a promising solution for
simulations of collisionless systems with large or arbitrary thermal motions.

The Vlasov simulation allows us to study the nonlinear effect of massive 
neutrinos during the LSS formation. Without being compromised by particle 
shot noise, our simulations accurately reproduce the observational signatures 
of massive neutrinos that are to be detected by ongoing and future wide-field 
galaxy surveys. The observations utilizing ground-based telescopes such as 
Vera C. Rubin Telescope and space-borne ones such as NASA's Nancy Grace Roman 
Telescope and ESA's Euclid will ultimately lead to precise determination of the 
absolute mass of neutrinos.

An array of state-of-the-art techniques are integrated to directly solve 
the six-dimensional Vlasov equation. Our novel advection scheme enables us to 
achieve spatially high-order (less diffusive) solutions with computationally 
light weight time integration. The whole implementation of this innovative 
scheme is highly optimized by exploiting SIMD instructions in the best possible manner.
To this end, we introduce a novel LAT method 
to pack regularly discretized data into SIMD registers efficiently.
The concerted use of the modern techniques and the SIMD instructions 
significantly reduces the total computational cost that is otherwise needed.

The parallel efficiency of our simulation is remarkably excellent for both the 
weak and strong scalings. This is partially in virtue of relatively 
monolithic, high-bandwidth and low-latency interconnect, the Tofu 
interconnect D, equipped with Fugaku supercomputer which directly 
connect sets of CMG and HBM2 memory embedded in a A64FX processor. 
Although stencil computations like Vlasov simulations are basically 
memory-bandwidth limited and may appear better suited to be performed on 
GPU-like processors, parallel efficiency on a massive parallel 
environment with such processors can be hampered by the 
multi-layered network stack. 
The very combination of our innovative numerical scheme,
highly optimized implementation and Fugaku supercomputer
achieves the high parallel efficiency and high computational 
performance simultaneously. 

The Vlasov equation~(\ref{eq:vlasov}) and the full Boltzmann equation
are classical first-principle equations that describe the collective and 
statistical behavior of many-particle systems in which the motion of particles 
is characterized by a certain Hamiltonian. Our scheme presented here can be 
applied to many other physical problems such as electrostatic and magnetized 
plasma phenomena and self-gravitating systems. 
Despite the huge computational cost, Vlasov simulations hold a clear
advantage that the velocity distribution function is represented as a
continuous function, and thus are well suited to simulate physical systems
where kinematic phenomena play an important role.
 
As an application of Vlasov simulations, one of the promising
targets would be numerical simulations of astrophysical magnetized
plasma such as interactions between inter-planetary plasma and planetary 
magnetospheres, and high energy plasma around astrophysical compact 
objects (black holes and neutron stars), in which a variety of kinematic 
phenomena such as particle acceleration induced  in collisionless shock waves, magneto-rotational instability and magnetic reconnection play 
critical roles in the dynamical evolution of these objects. Although particle-based Particle-In-Cell 
(PIC) simulations have a very successful history in this field, 
there exist several long standing and intrinsic difficulties
arising from the discreteness of numerical super-particles and the associated
shot noise. The Vlasov simulation of a magnetized plasma which integrate the Vlasov 
equation coupled with the Maxwell equations can be an interesting and 
straightforward extension of our approach.

In numerical cosmology, our hybrid approach consisting of Vlasov and particle-based 
$N$-body simulations places a milestone.
This approach takes the best advantage of both the particle-based 
and Vlasov simulations in a complementary manner, so that the simulated volume of our 
largest run covers a significant fraction of the entire observable Universe, 
while resolving nonlinear objects such as galaxy clusters.
The same approach can also be 
applied to plasma physics. For instance, the dynamics of heavy ions can be 
followed by a particle-based method whereas the electron dynamics is followed 
by the Vlasov simulation. We foresee that the hybrid approach opens a new 
paradigm in computational physics in the era of exa-scale supercomputing.

\begin{acks}
This research is supported by MEXT as “Priority Issue on post-K computer” 
(Elucidation of the Fundamental Laws and Evolution of the Universe) 
and “Program for Promoting Researches on the Supercomputer Fugaku” 
(Toward a unified view of the universe: from large scale structures to planets).
This research is also supported
by the JSPS KAKENHI Grant Number JP18H04336 and JP21H01079, by JST CREST JPMJCR1414 and by JST AIP Acceleration Research Grant JP20317829. Our code has been developed partially on ATERUI
supercomputer at Center for Computational Astrophysics (CfCA), National
Astronomical Observatory of Japan in its early stage. This research also used computational resources of the
Oakforest–PACS through the HPCI System Research Project (project
ID:hp170123 and hp190093) and Multidisciplinary Cooperative Research
Program in Center for Computational Sciences, University of Tsukuba (project ID:17a40 and xg18i019). 
\end{acks}

\bibliographystyle{ACM-Reference-Format}
\bibliography{ms}


\begin{thebibliography}{27}


\ifx \showCODEN    \undefined \def \showCODEN     #1{\unskip}     \fi
\ifx \showDOI      \undefined \def \showDOI       #1{#1}\fi
\ifx \showISBNx    \undefined \def \showISBNx     #1{\unskip}     \fi
\ifx \showISBNxiii \undefined \def \showISBNxiii  #1{\unskip}     \fi
\ifx \showISSN     \undefined \def \showISSN      #1{\unskip}     \fi
\ifx \showLCCN     \undefined \def \showLCCN      #1{\unskip}     \fi
\ifx \shownote     \undefined \def \shownote      #1{#1}          \fi
\ifx \showarticletitle \undefined \def \showarticletitle #1{#1}   \fi
\ifx \showURL      \undefined \def \showURL       {\relax}        \fi
\providecommand\bibfield[2]{#2}
\providecommand\bibinfo[2]{#2}
\providecommand\natexlab[1]{#1}
\providecommand\showeprint[2][]{arXiv:#2}

\bibitem[\protect\citeauthoryear{{Bagla}}{{Bagla}}{2002}]%
        {Bagla2002}
\bibfield{author}{\bibinfo{person}{J.~S. {Bagla}}.}
  \bibinfo{year}{2002}\natexlab{}.
\newblock \showarticletitle{{TreePM: A Code for Cosmological N-Body
  Simulations}}.
\newblock \bibinfo{journal}{\emph{Journal of Astrophysics and Astronomy}}
  \bibinfo{volume}{23} (\bibinfo{date}{Dec.} \bibinfo{year}{2002}),
  \bibinfo{pages}{185--196}.
\newblock
\urldef\tempurl%
\url{https://doi.org/10.1007/BF02702282}
\showDOI{\tempurl}
\showeprint[arxiv]{astro-ph/9911025}~[astro-ph]


\bibitem[\protect\citeauthoryear{{Banerjee}, {Powell}, {Abel}, and
  {Villaescusa-Navarro}}{{Banerjee} et~al\mbox{.}}{2018}]%
        {Banerjee2018}
\bibfield{author}{\bibinfo{person}{A. {Banerjee}}, \bibinfo{person}{D.
  {Powell}}, \bibinfo{person}{T. {Abel}}, {and} \bibinfo{person}{F.
  {Villaescusa-Navarro}}.} \bibinfo{year}{2018}\natexlab{}.
\newblock \showarticletitle{{Reducing noise in cosmological N-body simulations
  with neutrinos}}.
\newblock \bibinfo{journal}{\emph{\jcap}}  \bibinfo{volume}{9}, Article
  \bibinfo{articleno}{028} (\bibinfo{date}{Sept.} \bibinfo{year}{2018}),
  \bibinfo{numpages}{028}~pages.
\newblock
\urldef\tempurl%
\url{https://doi.org/10.1088/1475-7516/2018/09/028}
\showDOI{\tempurl}
\showeprint[arxiv]{1801.03906}


\bibitem[\protect\citeauthoryear{{Bird}, {Viel}, and {Haehnelt}}{{Bird}
  et~al\mbox{.}}{2012}]%
        {Bird2012}
\bibfield{author}{\bibinfo{person}{S. {Bird}}, \bibinfo{person}{M. {Viel}},
  {and} \bibinfo{person}{M.~G. {Haehnelt}}.} \bibinfo{year}{2012}\natexlab{}.
\newblock \showarticletitle{{Massive neutrinos and the non-linear matter power
  spectrum}}.
\newblock \bibinfo{journal}{\emph{\mnras}}  \bibinfo{volume}{420}
  (\bibinfo{date}{March} \bibinfo{year}{2012}), \bibinfo{pages}{2551--2561}.
\newblock
\urldef\tempurl%
\url{https://doi.org/10.1111/j.1365-2966.2011.20222.x}
\showDOI{\tempurl}
\showeprint[arxiv]{1109.4416}


\bibitem[\protect\citeauthoryear{{Cheng} and {Knorr}}{{Cheng} and
  {Knorr}}{1976}]%
        {Cheng1976}
\bibfield{author}{\bibinfo{person}{C.~Z. {Cheng}} {and} \bibinfo{person}{G.
  {Knorr}}.} \bibinfo{year}{1976}\natexlab{}.
\newblock \showarticletitle{{The Integration of the Vlasov Equation in
  Configuration Space}}.
\newblock \bibinfo{journal}{\emph{J. Comput. Phys.}} \bibinfo{volume}{22},
  \bibinfo{number}{3} (\bibinfo{date}{Nov} \bibinfo{year}{1976}),
  \bibinfo{pages}{330--351}.
\newblock
\urldef\tempurl%
\url{https://doi.org/10.1016/0021-9991(76)90053-X}
\showDOI{\tempurl}


\bibitem[\protect\citeauthoryear{{Cuperman}, {Harten}, and {Lecar}}{{Cuperman}
  et~al\mbox{.}}{1971}]%
        {Cuperman1971}
\bibfield{author}{\bibinfo{person}{S. {Cuperman}}, \bibinfo{person}{A.
  {Harten}}, {and} \bibinfo{person}{M. {Lecar}}.}
  \bibinfo{year}{1971}\natexlab{}.
\newblock \showarticletitle{{A Phase-Space Boundary Integration of the Vlasov
  Equation for Collisionless One-Dimensional Stellar Systems}}.
\newblock \bibinfo{journal}{\emph{\apss}} \bibinfo{volume}{13},
  \bibinfo{number}{2} (\bibinfo{date}{Oct} \bibinfo{year}{1971}),
  \bibinfo{pages}{411--424}.
\newblock
\urldef\tempurl%
\url{https://doi.org/10.1007/BF00649170}
\showDOI{\tempurl}


\bibitem[\protect\citeauthoryear{{Dubinski}, {Kim}, {Park}, and
  {Humble}}{{Dubinski} et~al\mbox{.}}{2004}]%
        {Dubinski2004}
\bibfield{author}{\bibinfo{person}{John {Dubinski}}, \bibinfo{person}{Juhan
  {Kim}}, \bibinfo{person}{Changbom {Park}}, {and} \bibinfo{person}{Robin
  {Humble}}.} \bibinfo{year}{2004}\natexlab{}.
\newblock \showarticletitle{{GOTPM: a parallel hybrid particle-mesh treecode}}.
\newblock \bibinfo{journal}{\emph{\na}} \bibinfo{volume}{9},
  \bibinfo{number}{2} (\bibinfo{date}{Feb.} \bibinfo{year}{2004}),
  \bibinfo{pages}{111--126}.
\newblock
\urldef\tempurl%
\url{https://doi.org/10.1016/j.newast.2003.08.002}
\showDOI{\tempurl}
\showeprint[arxiv]{astro-ph/0304467}~[astro-ph]


\bibitem[\protect\citeauthoryear{{Emberson}, {Yu}, {Inman}, {Zhang}, {Pen},
  {Harnois-D{\'e}raps}, {Yuan}, {Teng}, {Zhu}, {Chen}, and {Xing}}{{Emberson}
  et~al\mbox{.}}{2017}]%
        {Emberson2017}
\bibfield{author}{\bibinfo{person}{J.~D. {Emberson}}, \bibinfo{person}{Hao-Ran
  {Yu}}, \bibinfo{person}{Derek {Inman}}, \bibinfo{person}{Tong-Jie {Zhang}},
  \bibinfo{person}{Ue-Li {Pen}}, \bibinfo{person}{Joachim
  {Harnois-D{\'e}raps}}, \bibinfo{person}{Shuo {Yuan}},
  \bibinfo{person}{Huan-Yu {Teng}}, \bibinfo{person}{Hong-Ming {Zhu}},
  \bibinfo{person}{Xuelei {Chen}}, {and} \bibinfo{person}{Zhi-Zhong {Xing}}.}
  \bibinfo{year}{2017}\natexlab{}.
\newblock \showarticletitle{{Cosmological neutrino simulations at extreme
  scale}}.
\newblock \bibinfo{journal}{\emph{Research in Astronomy and Astrophysics}}
  \bibinfo{volume}{17}, \bibinfo{number}{8}, Article \bibinfo{articleno}{085}
  (\bibinfo{date}{Aug.} \bibinfo{year}{2017}), \bibinfo{numpages}{085}~pages.
\newblock
\urldef\tempurl%
\url{https://doi.org/10.1088/1674-4527/17/8/85}
\showDOI{\tempurl}
\showeprint[arxiv]{1611.01545}~[astro-ph.CO]


\bibitem[\protect\citeauthoryear{{Fujiwara}}{{Fujiwara}}{1981}]%
        {Fujiwara1981}
\bibfield{author}{\bibinfo{person}{T. {Fujiwara}}.}
  \bibinfo{year}{1981}\natexlab{}.
\newblock \showarticletitle{{Vlasov Simulations of Stellar Systems - Infinite
  Homogeneous Case}}.
\newblock \bibinfo{journal}{\emph{\pasj}}  \bibinfo{volume}{33}
  (\bibinfo{year}{1981}), \bibinfo{pages}{531}.
\newblock


\bibitem[\protect\citeauthoryear{{Fujiwara}}{{Fujiwara}}{1983}]%
        {Fujiwara1983b}
\bibfield{author}{\bibinfo{person}{T. {Fujiwara}}.}
  \bibinfo{year}{1983}\natexlab{}.
\newblock \showarticletitle{{Formation of Massive Galactic Halos with
  Neutrinos}}.
\newblock \bibinfo{journal}{\emph{Progress of Theoretical Physics}}
  \bibinfo{volume}{70} (\bibinfo{date}{Aug.} \bibinfo{year}{1983}),
  \bibinfo{pages}{603--605}.
\newblock
\urldef\tempurl%
\url{https://doi.org/10.1143/PTP.70.603}
\showDOI{\tempurl}


\bibitem[\protect\citeauthoryear{{Fukuda}, {Hayakawa}, {Ichihara}, {Inoue},
  {Ishihara}, {Ishino}, {Itow}, {Kajita}, {Kameda}, {Kasuga}, {Kobayashi},
  {Kobayashi}, {Koshio}, {Miura}, {Nakahata}, {Nakayama}, {Okada}, {Okumura},
  {Sakurai}, {Shiozawa}, {Suzuki}, {Takeuchi}, {Totsuka}, {Yamada}, {Earl},
  {Habig}, {Kearns}, {Messier}, {Scholberg}, {Stone}, {Sulak}, {Walter},
  {Goldhaber}, {Barszczxak}, {Casper}, {Gajewski}, {Halverson}, {Hsu}, {Kropp},
  {Price}, {Reines}, {Smy}, {Sobel}, {Vagins}, {Ganezer}, {Keig}, {Ellsworth},
  {Tasaka}, {Flanagan}, {Kibayashi}, {Learned}, {Matsuno}, {Stenger},
  {Takemori}, {Ishii}, {Kanzaki}, {Kobayashi}, {Mine}, {Nakamura}, {Nishikawa},
  {Oyama}, {Sakai}, {Sakuda}, {Sasaki}, {Echigo}, {Kohama}, {Suzuki}, {Haines},
  {Blaufuss}, {Kim}, {Sanford}, {Svoboda}, {Chen}, {Conner}, {Goodman},
  {Sullivan}, {Hill}, {Jung}, {Martens}, {Mauger}, {McGrew}, {Sharkey},
  {Viren}, {Yanagisawa}, {Doki}, {Miyano}, {Okazawa}, {Saji}, {Takahata},
  {Nagashima}, {Takita}, {Yamaguchi}, {Yoshida}, {Kim}, {Etoh}, {Fujita},
  {Hasegawa}, {Hasegawa}, {Hatakeyama}, {Iwamoto}, {Koga}, {Maruyama}, {Ogawa},
  {Shirai}, {Suzuki}, {Tsushima}, {Koshiba}, {Nemoto}, {Nishijima}, {Futagami},
  {Hayato}, {Kanaya}, {Kaneyuki}, {Watanabe}, {Kielczewska}, {Doyle}, {George},
  {Stachyra}, {Wai}, {Wilkes}, and {Young}}{{Fukuda} et~al\mbox{.}}{1998}]%
        {Fukuda1998}
\bibfield{author}{\bibinfo{person}{Y. {Fukuda}}, \bibinfo{person}{T.
  {Hayakawa}}, \bibinfo{person}{E. {Ichihara}}, \bibinfo{person}{K. {Inoue}},
  \bibinfo{person}{K. {Ishihara}}, \bibinfo{person}{H. {Ishino}},
  \bibinfo{person}{Y. {Itow}}, \bibinfo{person}{T. {Kajita}},
  \bibinfo{person}{J. {Kameda}}, \bibinfo{person}{S. {Kasuga}},
  \bibinfo{person}{K. {Kobayashi}}, \bibinfo{person}{Y. {Kobayashi}},
  \bibinfo{person}{Y. {Koshio}}, \bibinfo{person}{M. {Miura}},
  \bibinfo{person}{M. {Nakahata}}, \bibinfo{person}{S. {Nakayama}},
  \bibinfo{person}{A. {Okada}}, \bibinfo{person}{K. {Okumura}},
  \bibinfo{person}{N. {Sakurai}}, \bibinfo{person}{M. {Shiozawa}},
  \bibinfo{person}{Y. {Suzuki}}, \bibinfo{person}{Y. {Takeuchi}},
  \bibinfo{person}{Y. {Totsuka}}, \bibinfo{person}{S. {Yamada}},
  \bibinfo{person}{M. {Earl}}, \bibinfo{person}{A. {Habig}},
  \bibinfo{person}{E. {Kearns}}, \bibinfo{person}{M.~D. {Messier}},
  \bibinfo{person}{K. {Scholberg}}, \bibinfo{person}{J.~L. {Stone}},
  \bibinfo{person}{L.~R. {Sulak}}, \bibinfo{person}{C.~W. {Walter}},
  \bibinfo{person}{M. {Goldhaber}}, \bibinfo{person}{T. {Barszczxak}},
  \bibinfo{person}{D. {Casper}}, \bibinfo{person}{W. {Gajewski}},
  \bibinfo{person}{P.~G. {Halverson}}, \bibinfo{person}{J. {Hsu}},
  \bibinfo{person}{W.~R. {Kropp}}, \bibinfo{person}{L.~R. {Price}},
  \bibinfo{person}{F. {Reines}}, \bibinfo{person}{M. {Smy}},
  \bibinfo{person}{H.~W. {Sobel}}, \bibinfo{person}{M.~R. {Vagins}},
  \bibinfo{person}{K.~S. {Ganezer}}, \bibinfo{person}{W.~E. {Keig}},
  \bibinfo{person}{R.~W. {Ellsworth}}, \bibinfo{person}{S. {Tasaka}},
  \bibinfo{person}{J.~W. {Flanagan}}, \bibinfo{person}{A. {Kibayashi}},
  \bibinfo{person}{J.~G. {Learned}}, \bibinfo{person}{S. {Matsuno}},
  \bibinfo{person}{V.~J. {Stenger}}, \bibinfo{person}{D. {Takemori}},
  \bibinfo{person}{T. {Ishii}}, \bibinfo{person}{J. {Kanzaki}},
  \bibinfo{person}{T. {Kobayashi}}, \bibinfo{person}{S. {Mine}},
  \bibinfo{person}{K. {Nakamura}}, \bibinfo{person}{K. {Nishikawa}},
  \bibinfo{person}{Y. {Oyama}}, \bibinfo{person}{A. {Sakai}},
  \bibinfo{person}{M. {Sakuda}}, \bibinfo{person}{O. {Sasaki}},
  \bibinfo{person}{S. {Echigo}}, \bibinfo{person}{M. {Kohama}},
  \bibinfo{person}{A.~T. {Suzuki}}, \bibinfo{person}{T.~J. {Haines}},
  \bibinfo{person}{E. {Blaufuss}}, \bibinfo{person}{B.~K. {Kim}},
  \bibinfo{person}{R. {Sanford}}, \bibinfo{person}{R. {Svoboda}},
  \bibinfo{person}{M.~L. {Chen}}, \bibinfo{person}{Z. {Conner}},
  \bibinfo{person}{J.~A. {Goodman}}, \bibinfo{person}{G.~W. {Sullivan}},
  \bibinfo{person}{J. {Hill}}, \bibinfo{person}{C.~K. {Jung}},
  \bibinfo{person}{K. {Martens}}, \bibinfo{person}{C. {Mauger}},
  \bibinfo{person}{C. {McGrew}}, \bibinfo{person}{E. {Sharkey}},
  \bibinfo{person}{B. {Viren}}, \bibinfo{person}{C. {Yanagisawa}},
  \bibinfo{person}{W. {Doki}}, \bibinfo{person}{K. {Miyano}},
  \bibinfo{person}{H. {Okazawa}}, \bibinfo{person}{C. {Saji}},
  \bibinfo{person}{M. {Takahata}}, \bibinfo{person}{Y. {Nagashima}},
  \bibinfo{person}{M. {Takita}}, \bibinfo{person}{T. {Yamaguchi}},
  \bibinfo{person}{M. {Yoshida}}, \bibinfo{person}{S.~B. {Kim}},
  \bibinfo{person}{M. {Etoh}}, \bibinfo{person}{K. {Fujita}},
  \bibinfo{person}{A. {Hasegawa}}, \bibinfo{person}{T. {Hasegawa}},
  \bibinfo{person}{S. {Hatakeyama}}, \bibinfo{person}{T. {Iwamoto}},
  \bibinfo{person}{M. {Koga}}, \bibinfo{person}{T. {Maruyama}},
  \bibinfo{person}{H. {Ogawa}}, \bibinfo{person}{J. {Shirai}},
  \bibinfo{person}{A. {Suzuki}}, \bibinfo{person}{F. {Tsushima}},
  \bibinfo{person}{M. {Koshiba}}, \bibinfo{person}{M. {Nemoto}},
  \bibinfo{person}{K. {Nishijima}}, \bibinfo{person}{T. {Futagami}},
  \bibinfo{person}{Y. {Hayato}}, \bibinfo{person}{Y. {Kanaya}},
  \bibinfo{person}{K. {Kaneyuki}}, \bibinfo{person}{Y. {Watanabe}},
  \bibinfo{person}{D. {Kielczewska}}, \bibinfo{person}{R.~A. {Doyle}},
  \bibinfo{person}{J.~S. {George}}, \bibinfo{person}{A.~L. {Stachyra}},
  \bibinfo{person}{L.~L. {Wai}}, \bibinfo{person}{R.~J. {Wilkes}}, {and}
  \bibinfo{person}{K.~K. {Young}}.} \bibinfo{year}{1998}\natexlab{}.
\newblock \showarticletitle{{Evidence for Oscillation of Atmospheric
  Neutrinos}}.
\newblock \bibinfo{journal}{\emph{Physical Review Letters}}
  \bibinfo{volume}{81} (\bibinfo{date}{Aug.} \bibinfo{year}{1998}),
  \bibinfo{pages}{1562--1567}.
\newblock
\urldef\tempurl%
\url{https://doi.org/10.1103/PhysRevLett.81.1562}
\showDOI{\tempurl}
\showeprint{hep-ex/9807003}


\bibitem[\protect\citeauthoryear{{Hockney} and {Eastwood}}{{Hockney} and
  {Eastwood}}{1981}]%
        {Hockney1981}
\bibfield{author}{\bibinfo{person}{R.~W. {Hockney}} {and}
  \bibinfo{person}{J.~W. {Eastwood}}.} \bibinfo{year}{1981}\natexlab{}.
\newblock \bibinfo{booktitle}{\emph{{Computer Simulation Using Particles}}}.
\newblock \bibinfo{publisher}{McGraw-Hill}.
\newblock


\bibitem[\protect\citeauthoryear{Idomura, Ida, Kano, Aiba, and Tokuda}{Idomura
  et~al\mbox{.}}{2008}]%
        {Idomura2008}
\bibfield{author}{\bibinfo{person}{Yasuhiro Idomura}, \bibinfo{person}{Masato
  Ida}, \bibinfo{person}{Takuma Kano}, \bibinfo{person}{Nobuyuki Aiba}, {and}
  \bibinfo{person}{Shinji Tokuda}.} \bibinfo{year}{2008}\natexlab{}.
\newblock \showarticletitle{Conservative global gyrokinetic toroidal full-f
  five-dimensional Vlasov simulation}.
\newblock \bibinfo{journal}{\emph{Computer Physics Communications}}
  \bibinfo{volume}{179}, \bibinfo{number}{6} (\bibinfo{year}{2008}),
  \bibinfo{pages}{391--403}.
\newblock
\showISSN{0010-4655}
\urldef\tempurl%
\url{https://doi.org/10.1016/j.cpc.2008.04.005}
\showDOI{\tempurl}


\bibitem[\protect\citeauthoryear{{Inman}, {Emberson}, {Pen}, {Farchi}, {Yu},
  and {Harnois-D{\'e}raps}}{{Inman} et~al\mbox{.}}{2015}]%
        {Inman2015}
\bibfield{author}{\bibinfo{person}{D. {Inman}}, \bibinfo{person}{J.~D.
  {Emberson}}, \bibinfo{person}{U.-L. {Pen}}, \bibinfo{person}{A. {Farchi}},
  \bibinfo{person}{H.-R. {Yu}}, {and} \bibinfo{person}{J.
  {Harnois-D{\'e}raps}}.} \bibinfo{year}{2015}\natexlab{}.
\newblock \showarticletitle{{Precision reconstruction of the cold dark
  matter-neutrino relative velocity from N -body simulations}}.
\newblock \bibinfo{journal}{\emph{\prd}} \bibinfo{volume}{92},
  \bibinfo{number}{2}, Article \bibinfo{articleno}{023502}
  (\bibinfo{date}{July} \bibinfo{year}{2015}),
  \bibinfo{numpages}{023502}~pages.
\newblock
\urldef\tempurl%
\url{https://doi.org/10.1103/PhysRevD.92.023502}
\showDOI{\tempurl}
\showeprint[arxiv]{1503.07480}


\bibitem[\protect\citeauthoryear{{Inman}, {Yu}, {Zhu}, {Emberson}, {Pen},
  {Zhang}, {Yuan}, {Chen}, and {Xing}}{{Inman} et~al\mbox{.}}{2017}]%
        {Inman2017}
\bibfield{author}{\bibinfo{person}{D. {Inman}}, \bibinfo{person}{H.-R. {Yu}},
  \bibinfo{person}{H.-M. {Zhu}}, \bibinfo{person}{J.~D. {Emberson}},
  \bibinfo{person}{U.-L. {Pen}}, \bibinfo{person}{T.-J. {Zhang}},
  \bibinfo{person}{S. {Yuan}}, \bibinfo{person}{X. {Chen}}, {and}
  \bibinfo{person}{Z.-Z. {Xing}}.} \bibinfo{year}{2017}\natexlab{}.
\newblock \showarticletitle{{Simulating the cold dark matter-neutrino dipole
  with TianNu}}.
\newblock \bibinfo{journal}{\emph{\prd}} \bibinfo{volume}{95},
  \bibinfo{number}{8}, Article \bibinfo{articleno}{083518}
  (\bibinfo{date}{April} \bibinfo{year}{2017}),
  \bibinfo{numpages}{083518}~pages.
\newblock
\urldef\tempurl%
\url{https://doi.org/10.1103/PhysRevD.95.083518}
\showDOI{\tempurl}


\bibitem[\protect\citeauthoryear{{Janin}}{{Janin}}{1971}]%
        {Janin1971}
\bibfield{author}{\bibinfo{person}{G. {Janin}}.}
  \bibinfo{year}{1971}\natexlab{}.
\newblock \showarticletitle{{Numerical Experiments with a One-Dimensional
  Gravitational System by a Euler-Type Method}}.
\newblock \bibinfo{journal}{\emph{\aap}}  \bibinfo{volume}{11}
  (\bibinfo{date}{Mar} \bibinfo{year}{1971}), \bibinfo{pages}{188}.
\newblock


\bibitem[\protect\citeauthoryear{{Kawai}, {Fukushige}, {Makino}, and
  {Taiji}}{{Kawai} et~al\mbox{.}}{2000}]%
        {Kawai2000}
\bibfield{author}{\bibinfo{person}{Atsushi {Kawai}}, \bibinfo{person}{Toshiyuki
  {Fukushige}}, \bibinfo{person}{Junichiro {Makino}}, {and}
  \bibinfo{person}{Makoto {Taiji}}.} \bibinfo{year}{2000}\natexlab{}.
\newblock \showarticletitle{{GRAPE-5: A Special-Purpose Computer for N-Body
  Simulations}}.
\newblock \bibinfo{journal}{\emph{\pasj}}  \bibinfo{volume}{52}
  (\bibinfo{date}{Aug.} \bibinfo{year}{2000}), \bibinfo{pages}{659--676}.
\newblock
\urldef\tempurl%
\url{https://doi.org/10.1093/pasj/52.4.659}
\showDOI{\tempurl}
\showeprint[arxiv]{astro-ph/9909116}~[astro-ph]


\bibitem[\protect\citeauthoryear{{Nitadori}, {Makino}, and {Hut}}{{Nitadori}
  et~al\mbox{.}}{2006}]%
        {Nitadori2006}
\bibfield{author}{\bibinfo{person}{Keigo {Nitadori}},
  \bibinfo{person}{Junichiro {Makino}}, {and} \bibinfo{person}{Piet {Hut}}.}
  \bibinfo{year}{2006}\natexlab{}.
\newblock \showarticletitle{{Performance tuning of N-body codes on modern
  microprocessors: I. Direct integration with a hermite scheme on x86\_64
  architecture}}.
\newblock \bibinfo{journal}{\emph{\na}} \bibinfo{volume}{12},
  \bibinfo{number}{3} (\bibinfo{date}{Dec.} \bibinfo{year}{2006}),
  \bibinfo{pages}{169--181}.
\newblock
\urldef\tempurl%
\url{https://doi.org/10.1016/j.newast.2006.07.007}
\showDOI{\tempurl}
\showeprint[arxiv]{astro-ph/0511062}~[astro-ph]


\bibitem[\protect\citeauthoryear{{Planck Collaboration}, {Ade}, {Aghanim},
  {Arnaud}, {Ashdown}, {Aumont}, {Baccigalupi}, {Banday}, {Barreiro},
  {Bartlett}, and et~al.}{{Planck Collaboration} et~al\mbox{.}}{2016}]%
        {Planck2015XIII}
\bibfield{author}{\bibinfo{person}{{Planck Collaboration}},
  \bibinfo{person}{P.~A.~R. {Ade}}, \bibinfo{person}{N. {Aghanim}},
  \bibinfo{person}{M. {Arnaud}}, \bibinfo{person}{M. {Ashdown}},
  \bibinfo{person}{J. {Aumont}}, \bibinfo{person}{C. {Baccigalupi}},
  \bibinfo{person}{A.~J. {Banday}}, \bibinfo{person}{R.~B. {Barreiro}},
  \bibinfo{person}{J.~G. {Bartlett}}, {and} \bibinfo{person}{et al.}}
  \bibinfo{year}{2016}\natexlab{}.
\newblock \showarticletitle{{Planck 2015 results. XIII. Cosmological
  parameters}}.
\newblock \bibinfo{journal}{\emph{\aap}}  \bibinfo{volume}{594}, Article
  \bibinfo{articleno}{A13} (\bibinfo{date}{Sept.} \bibinfo{year}{2016}),
  \bibinfo{numpages}{A13}~pages.
\newblock
\urldef\tempurl%
\url{https://doi.org/10.1051/0004-6361/201525830}
\showDOI{\tempurl}
\showeprint[arxiv]{1502.01589}


\bibitem[\protect\citeauthoryear{{Qiu} and {Christlieb}}{{Qiu} and
  {Christlieb}}{2010}]%
        {Qiu2010}
\bibfield{author}{\bibinfo{person}{J.-M. {Qiu}} {and} \bibinfo{person}{A.
  {Christlieb}}.} \bibinfo{year}{2010}\natexlab{}.
\newblock \showarticletitle{{A conservative high order semi-Lagrangian WENO
  method for the Vlasov equation}}.
\newblock \bibinfo{journal}{\emph{J. Comput. Phys.}}  \bibinfo{volume}{229}
  (\bibinfo{date}{Feb.} \bibinfo{year}{2010}), \bibinfo{pages}{1130--1149}.
\newblock
\urldef\tempurl%
\url{https://doi.org/10.1016/j.jcp.2009.10.016}
\showDOI{\tempurl}


\bibitem[\protect\citeauthoryear{{Qiu} and {Shu}}{{Qiu} and {Shu}}{2011}]%
        {Qiu2011}
\bibfield{author}{\bibinfo{person}{J.-M. {Qiu}} {and} \bibinfo{person}{C.-W.
  {Shu}}.} \bibinfo{year}{2011}\natexlab{}.
\newblock \showarticletitle{{Conservative high order semi-Lagrangian finite
  difference WENO methods for advection in incompressible flow}}.
\newblock \bibinfo{journal}{\emph{J. Comput. Phys.}}  \bibinfo{volume}{230}
  (\bibinfo{date}{Feb.} \bibinfo{year}{2011}), \bibinfo{pages}{863--889}.
\newblock
\urldef\tempurl%
\url{https://doi.org/10.1016/j.jcp.2010.04.037}
\showDOI{\tempurl}


\bibitem[\protect\citeauthoryear{Shu and Osher}{Shu and Osher}{1988}]%
        {Shu1988}
\bibfield{author}{\bibinfo{person}{Chi-Wang Shu} {and} \bibinfo{person}{Stanley
  Osher}.} \bibinfo{year}{1988}\natexlab{}.
\newblock \showarticletitle{Efficient implementation of essentially
  non-oscillatory shock-capturing schemes}.
\newblock \bibinfo{journal}{\emph{J. Comput. Phys.}} \bibinfo{volume}{77},
  \bibinfo{number}{2} (\bibinfo{year}{1988}), \bibinfo{pages}{439 -- 471}.
\newblock
\showISSN{0021-9991}
\urldef\tempurl%
\url{https://doi.org/10.1016/0021-9991(88)90177-5}
\showDOI{\tempurl}


\bibitem[\protect\citeauthoryear{Suresh and Huynh}{Suresh and Huynh}{1997}]%
        {Suresh1997}
\bibfield{author}{\bibinfo{person}{A. Suresh} {and} \bibinfo{person}{H.T.
  Huynh}.} \bibinfo{year}{1997}\natexlab{}.
\newblock \showarticletitle{Accurate Monotonicity-Preserving Schemes with
  Runge–Kutta Time Stepping}.
\newblock \bibinfo{journal}{\emph{J. Comput. Phys.}} \bibinfo{volume}{136},
  \bibinfo{number}{1} (\bibinfo{year}{1997}), \bibinfo{pages}{83 -- 99}.
\newblock
\showISSN{0021-9991}
\urldef\tempurl%
\url{https://doi.org/10.1006/jcph.1997.5745}
\showDOI{\tempurl}


\bibitem[\protect\citeauthoryear{{Tanaka}, {Yoshikawa}, {Minoshima}, and
  {Yoshida}}{{Tanaka} et~al\mbox{.}}{2017}]%
        {Tanaka2017}
\bibfield{author}{\bibinfo{person}{Satoshi {Tanaka}}, \bibinfo{person}{Kohji
  {Yoshikawa}}, \bibinfo{person}{Takashi {Minoshima}}, {and}
  \bibinfo{person}{Naoki {Yoshida}}.} \bibinfo{year}{2017}\natexlab{}.
\newblock \showarticletitle{{Multidimensional Vlasov-Poisson Simulations with
  High-order Monotonicity- and Positivity-preserving Schemes}}.
\newblock \bibinfo{journal}{\emph{The Astrophysical Journal}}
  \bibinfo{volume}{849}, \bibinfo{number}{2}, Article \bibinfo{articleno}{76}
  (\bibinfo{date}{Nov} \bibinfo{year}{2017}), \bibinfo{numpages}{76}~pages.
\newblock
\urldef\tempurl%
\url{https://doi.org/10.3847/1538-4357/aa901f}
\showDOI{\tempurl}
\showeprint[arxiv]{1702.08521}~[physics.comp-ph]


\bibitem[\protect\citeauthoryear{{Tanikawa}, {Yoshikawa}, {Nitadori}, and
  {Okamoto}}{{Tanikawa} et~al\mbox{.}}{2013}]%
        {Tanikawa2013}
\bibfield{author}{\bibinfo{person}{Ataru {Tanikawa}}, \bibinfo{person}{Kohji
  {Yoshikawa}}, \bibinfo{person}{Keigo {Nitadori}}, {and}
  \bibinfo{person}{Takashi {Okamoto}}.} \bibinfo{year}{2013}\natexlab{}.
\newblock \showarticletitle{{Phantom-GRAPE: Numerical software library to
  accelerate collisionless N-body simulation with SIMD instruction set on x86
  architecture}}.
\newblock \bibinfo{journal}{\emph{\na}}  \bibinfo{volume}{19}
  (\bibinfo{date}{Feb.} \bibinfo{year}{2013}), \bibinfo{pages}{74--88}.
\newblock
\urldef\tempurl%
\url{https://doi.org/10.1016/j.newast.2012.08.009}
\showDOI{\tempurl}
\showeprint[arxiv]{1203.4037}~[astro-ph.IM]


\bibitem[\protect\citeauthoryear{Watanabe and Sugama}{Watanabe and
  Sugama}{2005}]%
        {Watanabe2005}
\bibfield{author}{\bibinfo{person}{T.-H Watanabe} {and} \bibinfo{person}{H
  Sugama}.} \bibinfo{year}{2005}\natexlab{}.
\newblock \showarticletitle{Velocity{\textendash}space structures of
  distribution function in toroidal ion temperature gradient turbulence}.
\newblock \bibinfo{journal}{\emph{Nuclear Fusion}} \bibinfo{volume}{46},
  \bibinfo{number}{1} (\bibinfo{date}{dec} \bibinfo{year}{2005}),
  \bibinfo{pages}{24--32}.
\newblock
\urldef\tempurl%
\url{https://doi.org/10.1088/0029-5515/46/1/003}
\showDOI{\tempurl}


\bibitem[\protect\citeauthoryear{{Yoshikawa}, {Yoshida}, and
  {Umemura}}{{Yoshikawa} et~al\mbox{.}}{2013}]%
        {Yoshikawa2013}
\bibfield{author}{\bibinfo{person}{K. {Yoshikawa}}, \bibinfo{person}{N.
  {Yoshida}}, {and} \bibinfo{person}{M. {Umemura}}.}
  \bibinfo{year}{2013}\natexlab{}.
\newblock \showarticletitle{{Direct Integration of the Collisionless Boltzmann
  Equation in Six-dimensional Phase Space: Self-gravitating Systems}}.
\newblock \bibinfo{journal}{\emph{The Astrophysical Journal}}
  \bibinfo{volume}{762}, Article \bibinfo{articleno}{116} (\bibinfo{date}{Jan.}
  \bibinfo{year}{2013}), \bibinfo{numpages}{116}~pages.
\newblock
\urldef\tempurl%
\url{https://doi.org/10.1088/0004-637X/762/2/116}
\showDOI{\tempurl}
\showeprint[arxiv]{1206.6152}~[astro-ph.IM]


\bibitem[\protect\citeauthoryear{{Yu}, {Emberson}, {Inman}, {Zhang}, {Pen},
  {Harnois-D{\'e}raps}, {Yuan}, {Teng}, {Zhu}, {Chen}, {Xing}, {Du}, {Zhang},
  {Lu}, and {Liao}}{{Yu} et~al\mbox{.}}{2017}]%
        {Yu2017}
\bibfield{author}{\bibinfo{person}{Hao-Ran {Yu}}, \bibinfo{person}{J.~D.
  {Emberson}}, \bibinfo{person}{Derek {Inman}}, \bibinfo{person}{Tong-Jie
  {Zhang}}, \bibinfo{person}{Ue-Li {Pen}}, \bibinfo{person}{Joachim
  {Harnois-D{\'e}raps}}, \bibinfo{person}{Shuo {Yuan}},
  \bibinfo{person}{Huan-Yu {Teng}}, \bibinfo{person}{Hong-Ming {Zhu}},
  \bibinfo{person}{Xuelei {Chen}}, \bibinfo{person}{Zhi-Zhong {Xing}},
  \bibinfo{person}{Yunfei {Du}}, \bibinfo{person}{Lilun {Zhang}},
  \bibinfo{person}{Yutong {Lu}}, {and} \bibinfo{person}{Xiangke {Liao}}.}
  \bibinfo{year}{2017}\natexlab{}.
\newblock \showarticletitle{{Differential neutrino condensation onto cosmic
  structure}}.
\newblock \bibinfo{journal}{\emph{Nature Astronomy}}  \bibinfo{volume}{1},
  Article \bibinfo{articleno}{0143} (\bibinfo{date}{Jul} \bibinfo{year}{2017}),
  \bibinfo{numpages}{0143}~pages.
\newblock
\urldef\tempurl%
\url{https://doi.org/10.1038/s41550-017-0143}
\showDOI{\tempurl}
\showeprint[arxiv]{1609.08968}~[astro-ph.CO]


\end{thebibliography}

\end{document}